\documentclass[12pt]{article} 
\usepackage{pdfpages}
\usepackage{lmodern}
\usepackage{graphicx} 
\usepackage[utf8]{inputenc}
\usepackage{color}
\usepackage{rotating}
\usepackage{adjustbox}
\usepackage{subcaption}
\usepackage{multirow} 
\usepackage{comment}
\usepackage{breakurl}
\usepackage{natbib}

\usepackage[T1]{fontenc} 
\usepackage{geometry}
\usepackage{amsmath}
\usepackage{etoolbox} 
\usepackage{amsthm}
\usepackage{amsfonts}
\usepackage{amssymb}
\usepackage[section]{placeins}
\usepackage{diagbox} 
\usepackage{hyperref}
\usepackage{bbm}
\usepackage{graphicx} 
\usepackage{setspace}
\makeatletter
\usepackage[symbol]{footmisc}

\begin{document}
	\singlespace
	
\begin{center}
	\textbf{Modelling the large and dynamically growing bipartite network of German patents and inventors} \\ 
	Cornelius Fritz$^1$\footnote[1]{Corresponding Author:  \href{mailto:cornelius.fritz@stat.uni-muenchen.de}{cornelius.fritz@stat.uni-muenchen.de}}, Giacomo De Nicola$^1$, Sevag Kevork$^1$,\\ Dietmar Harhoff$^2$, Göran Kauermann$^1$\hspace{.2cm}\\
	\vspace{0.3cm}
	$^1$Department of Statistics, LMU Munich, Germany \\
	$^2$Max Planck Institute for Innovation and Competition, Munich, Germany\hspace{.2cm}
\end{center}

	\begin{abstract}
We analyse the bipartite dynamic network of inventors and patents registered within the main area of electrical engineering in Germany to explore the driving forces behind innovation. The data at hand leads to a  bipartite network, where an edge between an inventor and a patent is present if the inventor is a co-owner of the respective patent. Since more than a hundred thousand patents were filed by similarly as many inventors during the observational period, this amounts to a massive bipartite network, too large to be analysed as a whole. Therefore, we decompose the bipartite network by utilising an essential characteristic of the network: most inventors tend to stay active only for a relatively short period, while new ones become active at each point in time. Consequently, the adjacency matrix carries several structural zeros. To accommodate for these, we propose a bipartite variant of the Temporal Exponential Random Graph Model (TERGM) in which we let the actor set vary over time, differentiate between inventors that already submitted patents and those that did not, and account for pairwise statistics of inventors. Our results corroborate the hypotheses that inventor characteristics and knowledge flows play a crucial role in the dynamics of invention.
	 

	\end{abstract}
	\section{Introduction}
	\FloatBarrier

\label{sec:introduction}

In the social sciences, bipartite networks are often used to represent and study affiliation of the actors to some groups (such as directors on boards, \citealp{friel2016}, or football players in teams, \citealp{onody2004}) and participation of people to events (such as researchers citing papers, \citealp{small1973}, or actors in movies, \citealp{ahmed2007}). Research on bipartite structures was initially focused on unimodal projections of the networks \citep{Breiger1974}, where we consider two nodes of one type to be tied if they share at least one alter of the other kind. This practice forces the researcher to give priority to one type of node over another and thus comes with a loss of possibly relevant information \citep{koskinen2012}. Direct bipartite network analysis has first been considered in \citet{borgatti1997network}, where an introduction and traditional network analysis techniques are systematically discussed. \citet{latapy2008basic} further adjusted known concepts from unipartite networks, such as {clustering} and {redundancy}, to the bipartite case, with a focus on large networks. 

\begin{figure}[t!]
    \centering
    \includegraphics[width=0.7\textwidth, page = 2]{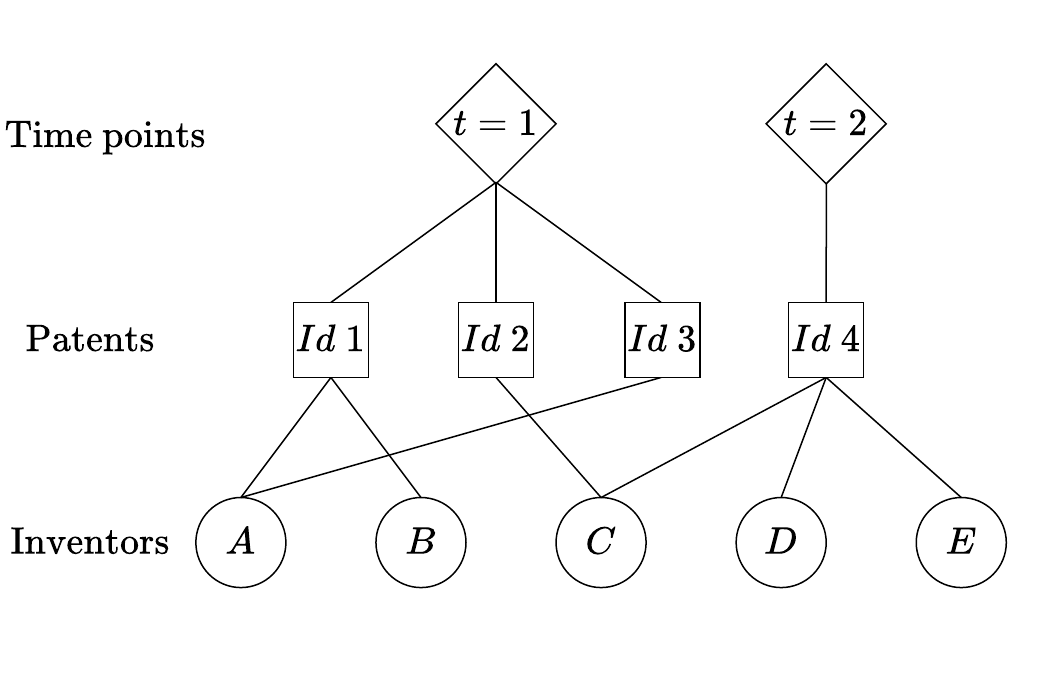}
    \caption{Illustrative example of the network structure.}
    \label{fig:alea1}
\end{figure}

For this paper, we also consider high-dimensional bipartite networks where actors are related to one another through instantaneous events, which by definition only occur once. In particular, we focus on the network formed by inventors residing in Germany and patents submitted between 1995 and 2015, where a tie between an inventor and a patent is present if the focal individual is listed among the patent's inventors. The resulting data structure is visualised in Figure \ref{fig:alea1}, where we can assign each patent (or event, in the jargon of bipartite network analysis)  to a time point and a set of co-inventors. 
For instance, inventors A and B filed the joint patent with ID 1. We may represent the bipartite network structure as an adjacency matrix with entries $Y_{ij}$, where 
\begin{equation}
\label{eq:matrix}
Y_{ij}  = 
\left\{ 
\begin{array}{ll} 
1 & \mbox{ if actor $i$ is on patent ID $j$ }\\
0 & \mbox{otherwise}
\end{array}  
\right.
\end{equation}
and $i \in \mathcal{I}$ and $j\in \mathcal{K}$, where we denote the complete set of inventors and patents by $\mathcal{I}$ and $\mathcal{K}$, respectively. In our example this bipartite network is of massive dimensions,
with $|\mathcal{I}|$ = 78.412 inventors on a total of $|\mathcal{K}|$= 126.388 filed patents. 


\begin{figure}[t!]
    \centering
    \includegraphics[width=0.65\textwidth, page = 2]{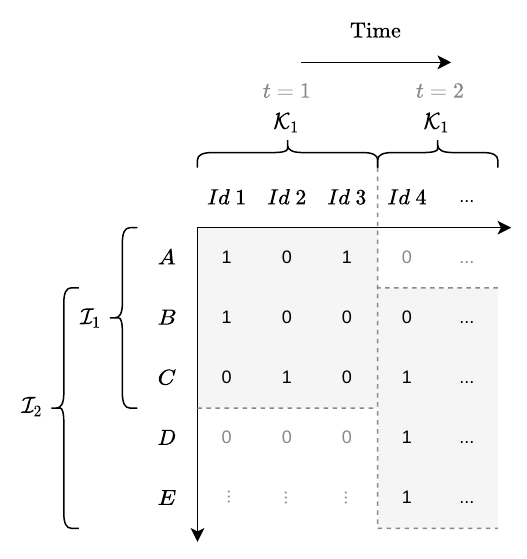}
    \caption{Illustrative example of the tripartite network matrix. The two sub-matrices shaded in grey are $\mathbf{Y}_1$  in the top left and $\mathbf{Y}_2$ on the bottom right. }
    \label{fig:alea2}
\end{figure}

The data allow us to gain insight into the dynamics and drivers of innovation, collaboration and knowledge flows in the private sector. Moreover, inventorship status on a patent is legally more binding than authorship of academic papers, suggesting a greater degree of validity of the results of network analysis in this context. The data, however, present some obstacles to their study. First, the complete network is too massive, making analysis with most traditional network techniques prohibitive. Second, the data carry structural zero entries since not all inventors are active during the entire time period between 1995 and 2015. This phenomenon is partially due to the retirement of inventors, who hence have natural ``{actor mortality}''. Concurrently, inventors are often active for a short period before changing careers, thereby ending their patent output and further reinforcing the aforementioned actor mortality. Vice versa, new inventors continuously enter the market by producing their first patent, resulting in what we can call ``{actor natality}'' in the network. These aspects imply that the bipartite network matrix (\ref{eq:matrix}) contains structural zeros for inventors which are not active at particular time points. 
To incorporate this feature into a statistical network model, we consider the network dynamically and discretise the time dimension by looking at yearly data, such that time takes values $t = 1,2, \ldots, T$, as sketched in Figure \ref{fig:alea1}. In this context, $T$ denote the number of observed time points. We then allow the actor set to change at each time point. For the adjacency matrix of Figure \ref{fig:alea1}, this leads to the matrix structure in Figure \ref{fig:alea2}, where e.g.\  inventor $A$ retires after time point $t = 1$ and hence does not take part in the patent market at $t=2$. We, therefore, define activity sets ${\cal I}_t$ to include all actors that are active at time point $t$. We also define the event set  ${\cal K}_t$, which contains all patents submitted in a particular time window. We assume that both sets are known for each time point $t = 1 , \ldots, T$. Therefore, we decompose the observed massive bipartite network matrix into smaller dimensional bipartite submatrices denoted by
\begin{align}
\mathbf{Y}_t = (Y_{ij} : i \in {\mathcal I}_t, j \in {\mathcal K}_t),
\end{align}
which are visualised for $t = 1$ and $2$  by the grey-shaded areas in  Figure \ref{fig:alea2}. Instead of modelling the entire bipartite network, we break down our analysis to 
modelling $\mathbf{Y}_t$ given the previous bipartite networks $\mathbf{Y}_1, \ldots \mathbf{Y}_{t-1}$. 
Incorporating the varying actor set as such in the analysis allows to structurally account for the observed actor mortality and natality while also making the difficulty of the problem more manageable, thus solving both issues simultaneously.

This change in perspective induces a structure that deviates from conventionally analysed networks. To accommodate for it in a probabilistic modelling framework, we extend the Temporal Exponential Random Graph Model (TERGM, \citealp{hanneke2010discrete}) towards dynamic bipartite networks with varying actor sets.
For TERGMs, we assume that a discrete Markov chain can describe the generating process of the networks observed over time. The transition probabilities of jumping from one network to another one are determined by an Exponential Random Graph Model (ERGM,\citealp{wasserman1996}). 
ERGMs, on the other hand, were adapted to bipartite data by \citet{Faust1999}, while adjustments to incorporate the model specifications of \citet{snijders2006} were proposed in \citet{wang2013exponential}.
These types of network models were already successfully applied to static \citep{agneessens2008local} as well as dynamic networks \citep{Broekel2018}. 

In addition to the dynamically varying actor set, the network at hand presents another particular feature for which we need to account in the modelling. Collaborations and knowledge flows generally build up over time, rather than being confined to single time points. To adequately represent these mechanisms, we need to include covariate information from the past and on the pairwise level of one alter set in the model, which has not yet been implemented in the bipartite ERGM framework. We, therefore, define and include novel sufficient network statistics in our model to account for this particular kind of dynamic interdependence. 

Overall, the contributions of this paper are the following. First, we demonstrate how massive bipartite networks can be broken down in a way that allows their analysis. Secondly, we extend temporal network models towards bipartite network data with varying actor sets. And lastly, with these requisites, we can analyse patent data with respect to innovation dynamics and collaboration in a more refined way than has been feasible to date. 

The remainder of the paper is organised as follows: Section \ref{sec:patent_data} gives a literature overview of the research in patent data. In this section, we also describe the data in detail. Section \ref{sec:model_spec_est} motivates the model and introduces its novelties in more detail from a theoretical perspective. We present the results of our empirical analysis in Section \ref{sec:application}, while Section \ref{sec:discussion} wraps up the paper with some concluding remarks.

\section{Patent data}
\label{sec:patent_data}
\subsection{Research in patent data} The analysis of patents and their impact and evolution over time is an important area of current economic research. \citet{Hall2011} provide a general overview of the field and its recent developments. The existence of patents induces a trade-off for society, namely between short-term monopoly rights to the use of inventions as an incentive to invent and early publication of inventions (rather than their secrecy for personal gain by the inventors), which may invite others to build on the patented technology. The study of patents in much of classic economic literature revolves around the economic consequences of their existence from a regulatory standpoint, or, in other words, whether the aforementioned trade-off is worth it for society. Patent data are often also used in innovation research to explore how new technologies develop and spread, which innovation areas are the most active, how innovation areas and sub-areas are connected with one another, and how productive firms or nations are with regards to their patenting output. 

Two of the main categories of methods for analysing patent data in this latter regard are keyword-based morphological approaches and network-based approaches. Under keyword-based patent analysis, we understand the process of gaining insights on core technology information of patents through text mining of the document content of each patent. \citet{Tseng2007} review this approach and describe different text mining techniques that conform to the analytical process used by patent analysts, see, e.g., \citet{Yoon2007} or \citet{Lee2009}.

In contrast to text mining approaches, network-based patent analysis starts with the idea of constructing a network map of the technology space to understand how technology as a whole behaves and evolves. 
In this context, patents (or technologies) are often considered as nodes of the network, where an edge links them if they are close to one another according to some proximity measure. Such measure is often based on the citations that occur between patents \citep{Alstott2017}. The study of citation networks has generally been an important area of research at least since the work of \citet{garfield1955} (see also \citealp{price1965} and \citealp{egghe1990}), and the techniques developed for general citation networks can naturally be applied to map patent citation networks (see, e.g., \citealp{Vonwartburg2005, Li2007, Verspagen2012}).

As an alternative to using citation-based approaches for measuring proximity, one can draw a network map of the technology space by focusing on the co-inventorship of patents. Using inventorship data instead of citations to construct the network entails a different perspective on the technological space: The focus is shifted from the content of the patents towards the people coming up with inventions. Thereby, we can gain insights into the network of inventors' underlying collaborative structure and investigate how behaviour differs between and among areas. 
Co-authorship networks have been extensively studied within the area of research publications (see, e.g., \citealp{melin1996} and \citealp{newman2004}). For patent data, it is possible to construct the co-inventorship network in two main ways. One can directly analyse the bipartite network formed by the patents and their inventors, see, e.g., \citet{Balconi2004}. Alternatively, one projects the bipartite structure on one of the two modes, which in the context of patent data is usually that of inventors. This entails a network composed only of inventors, in which two nodes are connected if they have at least one patent in common \citep{Ejermo2006,Bauer2021}. Much of the literature in this area utilises such projection since the focus is generally on knowledge flows between inventors, and models for unimodal networks are developed to a greater extent. As explained in the introduction, however, projecting everything on one mode inevitably loses information on the mode that is excluded.

\subsection{Data description}

\begin{figure}[t!]
    \centering
    \includegraphics[width=0.75\textwidth]{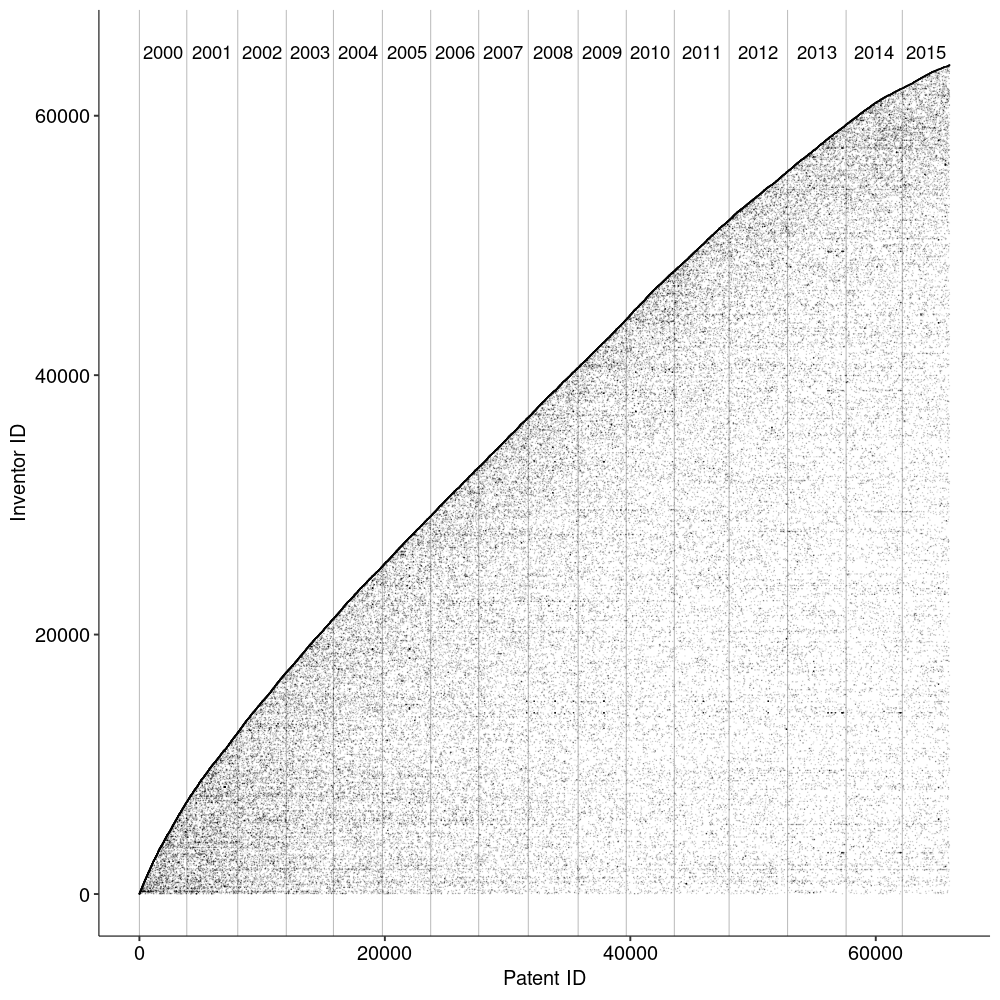}
    \caption{Adjacency matrix of the patent-inventor network between 2000 and 2015 with black points in the $(i,k)$th row indicating that inventor $i$ is a co-owner of patent $k$. Note that the points are heavily over-represented pixel-wise, hence the network is more sparse than it appears in the plot.}
    \label{fig:hugematrix}
\end{figure}

We consider patent applications submitted to the European Patent Office or the German Patent and Trademark Office (Deutsches Patent- und Markenamt) between 1995 and 2015. More specifically, we look at patents filed within the main area of electrical engineering, and for which at least one of the inventors listed on the patent has a residential address in Germany. For assigning each patent to a single time point we use the priority date, i.e., the first-time filing date of a patent (which precedes the publication and the grant date). We focus on electrical engineering as it is one of the largest main areas and as it has seen particularly high growth rates since 2010. Moreover, collaborations between inventors are particularly frequent in this field.  For our analyses, we focus on the data starting 2000 and condition on the information from the first five years considered (i.e. from 1995 to 1999) to derive covariates from them. The dataset can be represented in a massive bipartite network, for which the observed adjacency matrix (\ref{eq:matrix}) is visualised in Figure \ref{fig:hugematrix}. 

\begin{figure}[t!]
    \centering
    \includegraphics[width=0.6\textwidth, page = 1]{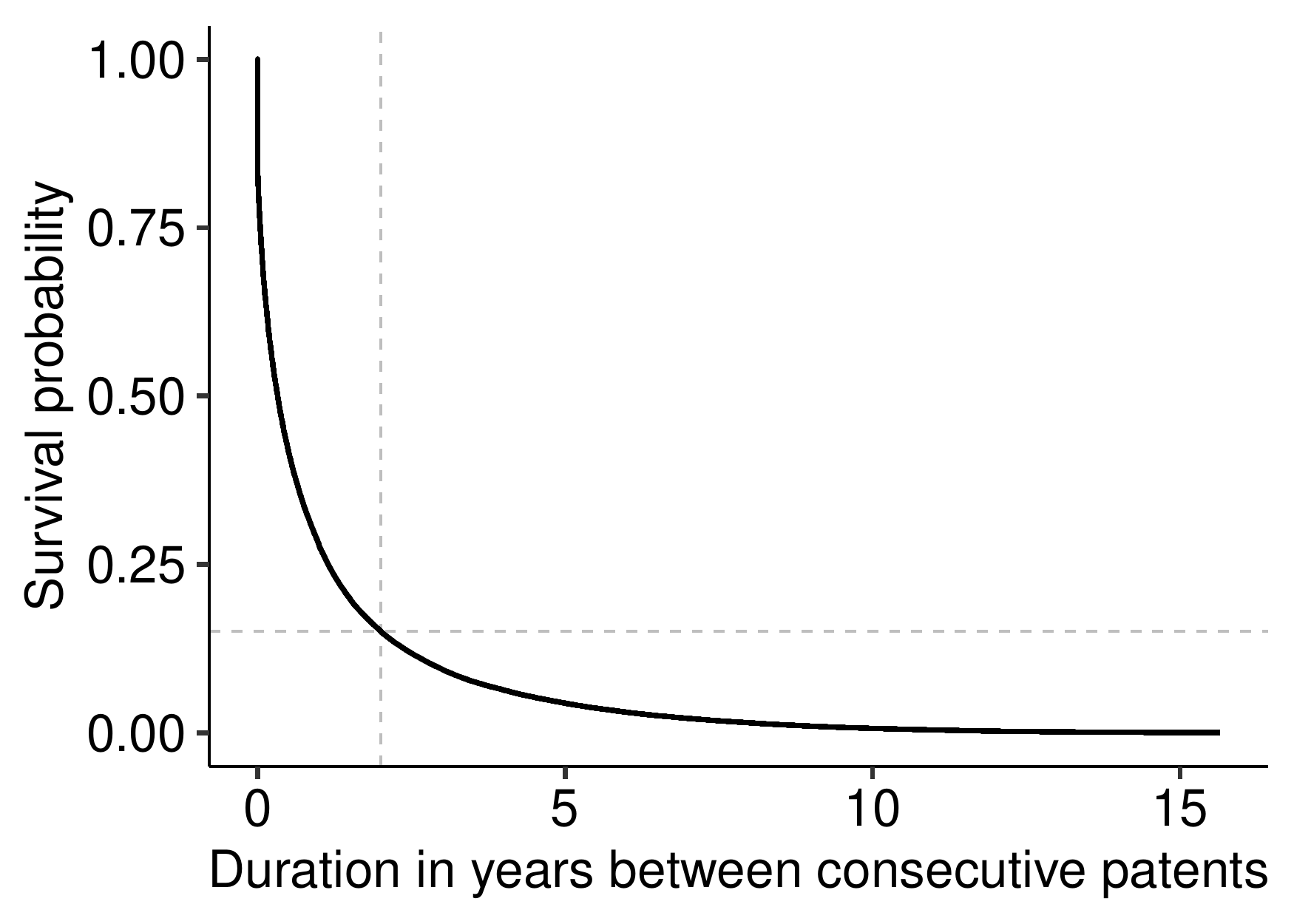}
    \caption{Kaplan-Meier estimate of the duration between consecutive patents submitted between 2000 and 2015. }
    \label{fig:km}
\end{figure}

As described in Section \ref{sec:introduction}, we instead consider this a dynamic bipartite network, discretising the time steps yearly such that time takes values $t = 1,2, \ldots, T$. In our notation, $t = 1$ translates to the year 2000. We also allow the actor set to change at each time point so that we end up with $T$ bipartite networks in which the nodes are given by the active inventors at each time point. 
Resulting from this, we include new inventors that are active for the first time and remove inactive ones from the network at each time point $t$. The latter point is motivated by the empirical data, which suggests that if previously active inventors don't produce any patents for a long time, it is likely that they will not be active anymore. This phenomenon can stem from a changed career path (moved up to a management position where writing patents is not among the work tasks) or retirement. To this point, we show in Figure \ref{fig:km} the Kaplan-Meier estimate of the time passing between two consecutive patents by the same inventor. As indicated by the dashed grey lines, about $85\%$ of patents by a specific inventor that already had at least one patent are submitted within two years from the previous one. Given this, we define an inventor as active at time $t$ if they had at least one patent in the two years prior to $t$. Note that by doing so we do not disregard the remaining $15\%$ of the data, but simply label these inventors as inactive for a specific period, at least until they appear on another patent.

As we are interested in investigating the drivers of patented innovation and inventor collaboration, we exclude patents developed by a single inventor from the modelled patent set. Moreover, we exclude inventors with no address in Germany from the actor set, as they make up less than 1\% of the population.
In addition to the residence address of each inventor and the date of each patent, we also incorporate information on the gender of each inventor in our model. 

	\FloatBarrier

\section{Modelling patent data as bipartite networks}
	\FloatBarrier

\label{sec:model_spec_est}
\subsection{Temporal Exponential Random Graph Models for bipartite networks}
\label{sec:tergm}
Having laid out the available data, we now formulate a generative network model for the bipartite networks at hand. This framework should allow us to differentiate between random and structural characteristics of the network to support or disregard our substantive expectations, such as, for example, whether or not two inventors that teamed up in the past are likely to produce another patent together in the future.
To do so we first need to introduce some additional notation. As a general rule, we write  $\mathbf{Y}_{t}$ for the network when viewed as a random variable and $\mathbf{y}_{t}= (y_{t,ik}: i \in \mathcal{I}_t, k \in \mathcal{K}_t)$ if we relate to the observed counterpart.  In this context, $y_{t,ik} = 1$ translates to inventor $i$ being a co-owner of patent $k$, while $y_{t,ik} = 0$ indicates the contrary. As a result, the observed networks are binary and undirected, i.e. $\mathbf{y}_{t} \in \{0,1\}^{|\mathcal{I}_t| \times |\mathcal{K}_t|}$. We denote the space of all networks that could potentially be observed at time point $t$ by $\mathcal{Y}_{t}$. For our application, as explained in the previous section, the latter is restricted to only allow for patents which have at least two inventors. 

We specify the joint probability for the set of networks through 
\begin{align}
\label{eq:joint_probability}
\mathbb{P}_{\boldsymbol{\theta}}(\mathbf{Y}_1, ..., \mathbf{Y}_T) =  \prod_{t = 1}^T\mathbb{P}_{\boldsymbol{\theta}}(\mathbf{Y}_t|\mathcal{H}_{t}),
\end{align}
where $\mathcal{H}_{t} $ defines the history, composed of previous bipartite networks 
$\mathbf{y}_1, ..., \mathbf{y}_{t-1}$ and covariates $\textbf{x}_1, ..., \textbf{x}_{t-1}$. 
The covariates can encompass dyadic and nodal information, but to make the notation less cumbersome, we suppress the explicit inclusion of the covariates in the formulae. 
We simplify \eqref{eq:joint_probability} by assuming a fixed time lag, i.e. 
\begin{align}
\label{eq:cond_probability}
\mathbb{P}_{\boldsymbol{\theta}}(\mathbf{Y}_t|\mathcal{H}_{t}) =  \mathbb{P}_{\boldsymbol{\theta}}(\mathbf{Y}_t|\mathbf{Y}_{t-1} = \mathbf{y}_{t-1}, ... , \mathbf{Y}_{t-s} =  \mathbf{y}_{t-s} ),
\end{align}
for $s\in \mathbb{N}$. The Markov property then allows us to postulate an ERGM for the transition probability \eqref{eq:cond_probability} in the following form:
\begin{align}\label{eq:tergm}
\mathbb{P}_{\boldsymbol{\theta}}(\mathbf{Y}_t|\mathbf{Y}_{t-1}, \mathbf{y}_{t-1}, ... , \mathbf{Y}_{t-s} =  \mathbf{y}_{t-s} ) = \frac{\exp \left\{ \boldsymbol{\theta}^\top\boldsymbol{s}( \mathbf{y}_{t},..., \mathbf{y}_{t-s}) \right\}}{\kappa(\boldsymbol{\theta}, \mathbf{y}_{t-1}, ..., \mathbf{y}_{t-s})},
\end{align}
where $\boldsymbol{\theta} = (\theta_1, \ldots , \theta_q)  \in \mathbb{R}^{q}$ is a $q$-dimensional vector of parameters, $\boldsymbol{s}():\mathcal{Y}_{t} \times ... \times \mathcal{Y}_{t-s} \rightarrow \mathbb{R}^q$ is the vector of sufficient statistics and \newline $\kappa(\boldsymbol{\theta}, \mathbf{y}_{t-1}, ..., \mathbf{y}_{t-s}) := \sum_{y \in \mathcal{Y}_t}\exp  \left\{ \boldsymbol{\theta^{\top}}\boldsymbol{s}(\mathbf{y}, \mathbf{y}_{t-1}, ..., \mathbf{y}_{t-s})\right\}$ is a normalising factor. We obtain a canonical exponential family model with known characteristics \citep{Barndorff-Nielsen1978}, which come in handy when quantifying the uncertainty of the estimates of $\boldsymbol{\theta}$. 
Note that for the application to patent data, the coefficients governing the transition from one time point to another are not necessarily constant over time due to external shocks, such as for example the dot-com bubble and the 2008 financial crisis, which may affect the activity of inventors. For this reason, we let $\boldsymbol{\theta}$ in (\ref{eq:tergm}) flexibly depend on time, and estimate it separately for each time point $t$. We omit the subscript $t$ from the formulae for notational simplicity though.

Interpreting the coefficients $\boldsymbol{\theta}$ can be done both at the global network level as well on the single tie level. For the former, $\theta_p>0$ implies that the expected value of the $p$th statistic of $\boldsymbol{s}( \mathbf{y}_{t},..., \mathbf{y}_{t-s})$  for networks generated from \eqref{eq:tergm} is higher than under a Bernoulli graph, while $\theta_p<0$ implies that it is lower. In this context, a Bernoulli graph is a simplistic network model where every edge is present with equal probability of 0.5.  For the latter, we define so called {change statistics}, which are the change in the sufficient statistics caused by switching the entry $y_{t,ik}$ from 0 to 1. Formally, 
\begin{align}
\label{eq:change_statistics}
\boldsymbol{\Delta}_{ik}(\mathbf{y}_{t},..., \mathbf{y}_{t-s}) = \boldsymbol{s}(\mathbf{y}_{t,ik}^+,..., \mathbf{y}_{t-s}) - \boldsymbol{s}(\mathbf{y}_{t,ik}^-,..., \mathbf{y}_{t-s}),
\end{align}
where $\mathbf{y}_{t,ik}^+$ is the network $\mathbf{y}_{t}$ with entry $y_{t,ik}$ fixed at 1, while the entry is set to 0 in  $\mathbf{y}_{t,ik}^-$. 
For each possible inventor-patent connection, we can then state the corresponding probability conditionally on the remaining bipartite network denoted by $\mathbf{y}_{t,ik}^C$, i.e.\ the complete network $\mathbf{y}_{t}$ excluding the single entry $y_{t,ik}$
. 
This leads to  
\begin{align}
\label{eq:change_stat}
\mathbb{P}_{\boldsymbol{\theta}}\left(Y_{t,ik} = 1| \mathbf{Y}_{t,ik}^C = \mathbf{y}_{t,ik}^C\right) = \frac{\exp\{\boldsymbol{\theta}^\top  \boldsymbol{\Delta}_{t,ik}(\mathbf{y}_{t},..., \mathbf{y}_{t-s})\}}{1 + \exp\{\boldsymbol{\theta}^\top  \boldsymbol{\Delta}_{t,ik}(\mathbf{y}_{t},..., \mathbf{y}_{t-s})\}}. 
\end{align}
Through this expression we can relate $\boldsymbol{\theta}$, the canonical parameter of \eqref{eq:tergm}, to the conditional probability of inventor $i$ to be co-owner of patent $k$. We can thereby derive an interpretation of the coefficients reminiscent of the common logistic regression: if adding the tie $y_{t,ik}$ to the network raises the $p$th entry of $\boldsymbol{\Delta}_{t,ik}(y_t,..., y_{t-s})$ by one unit, the conditional log-odds of $Y_{t,ik}$ are {ceteris paribus} altered by the additive factor $\theta_p$ \citep{StevenM2009}. 

\subsection{Sufficient statistics for bipartite patent data}
\label{sec:sufficient}
	\FloatBarrier

The main ingredient of model \eqref{eq:tergm} is the set of sufficient statistics, which translates to a particular dependence structure assumed for the edges in the observed bipartite network \citep{Wang2013}. A statistic that is typically included is the number of edges at time point $t$, i.e. $s_{\mathtt{edges}}(\mathbf{y}_{t}, ..., \mathbf{y}_{t-s}) = |\mathbf{y}_{t}|$, which can be comprehended as the equivalent of an intercept term in standard regression models \citep{StevenM2009}. 
As we are in a dynamic setting in which additional information on past networks is available, we can define statistics that depend on the past networks, such as the number of patents in the previous $s$ years for each actor active at time point $t$: 
\begin{align}
    \label{eq:stat_past_patents}
    s_{\mathtt{pastpatent}}(\mathbf{y}_{t}, ..., \mathbf{y}_{t-s}) = \sum_{i \in \mathcal{I}_t} \sum_{k \in \mathcal{K}_t} y_{t,ik} \sum_{u = t-s}^{t-1} \sum_{l \in \mathcal{K}_u} y_{u,il}
\end{align}
As the patent network presents some particular dependence structures, novel types of statistics are needed, which we describe in the following. 

\subsubsection{Pairwise statistics of inventors}


One drawback of representing the patent data as a bipartite adjacency matrix instead of the one-mode-projected version is that incorporating information on the pairwise inventor-to-inventor level is not straightforward. 
We therefore introduce assortative two-star statistics extending the work of \citet[Chapter 2]{Bomiriya2014} on homophily, which is defined as the mechanism driving ties between similar individuals \citep{McPherson2001}, for bipartite networks. 
We take the  patent-based two-star statistic as starting point, which for $\mathbf{y}_t$ is defined by
\begin{align}
\label{eq:2star}
    s_{\mathtt{twostar.patent}}(\mathbf{y}_t) = \frac{1}{2} 
    \sum_{k \in \mathcal{K}_t} \sum_{i \in \mathcal{I}_t} 
    y_{t,ik} \left(\sum_{j\neq i} y_{t,jk}\right).
\end{align}
The tendency to interact with one another is often based on the similarity of a factor variable $u_t = (u_{t,i}; i \in \mathcal{I}_t)$ 
. We therefore define the indicator matrix $\mathbf{x}_t \in \{0,1\}^{{|\cal I}_t|\times |\mathcal{I}_t|}$ with entries $x_{t,ij} = \mathbbm{1}(u_{t,i} = u_{t,j})$. In line with \citet[Chapter 2]{Bomiriya2014}, this allows to augment the two-star statistic \eqref{eq:2star} in the form
\begin{align}
\label{eq:2star_homo}
    s_{\mathtt{homophily.x}}(\mathbf{y}_t) = \frac{1}{2} \sum_{k \in \mathcal{K}_t} \sum_{i \in \mathcal{I}_t} y_{t,ik} \left(\sum_{j\neq i} y_{t,jk} x_{t,ij}\right).
\end{align}
Next, we generalise \eqref{eq:2star_homo} by not restricting ourselves to any particular definition of $\mathbf{x}_t$ but letting the matrix be an arbitrary function of the networks from the past $s$ years and other exogenous information. To further correct for different sizes of patents, i.e. the number of inventors co-owning the patent, we normalise the statistic by the degree of each patent, whereby the novel statistic is defined through: 
\begin{align}
\label{eq:2star_assortative}
    s_{\mathtt{assort.x}}(\mathbf{y}_t, ..., \mathbf{y}_{t-s}) =\frac{1}{2} \sum_{k \in \mathcal{K}_t} \sum_{i \in \mathcal{I}_t} y_{ik} \left(100\times \frac{ \sum_{j\neq i} y_{t,jk}x_{t,ij}}{    \sum_{j\neq i} y_{t,jk}}\right)
\end{align}
The corresponding change statistic for an edge between inventor $i$ and patent $k$ is  then
\begin{align}
\label{eq:change_stat_2star_assortative}
    \boldsymbol{\Delta}_{ik,\mathtt{assort.x}}(\mathbf{y}_t, ..., \mathbf{y}_{t-s}) = 100\times \frac{ \sum_{j\neq i} y_{t,jk}x_{t,ij}}{    \sum_{j\neq i} y_{t,jk}},
\end{align}
which can simply be interpreted as the percentage of inventors on patent $k$ that match with inventor $i$ in matrix $\mathbf{x}$. We multiply the statistic by 100, which does not affect the model itself but eases interpretation (as a unit increase is now equivalent to a single percentage change). To give an example of a statistic of this type, we can combine \eqref{eq:change_stat_2star_assortative} with matrix $\mathbf{x}_{t}^{P}$, for which entry $x_{t,ij}^P$ is 1 if inventor $i$ and $j$ already had a joint patent in the last $s$ years and 0 otherwise. The resulting statistic measures how previous collaboration among inventors affects the propensity of future collaboration. More examples for such statistics are provided in Section \ref{sec:application}. 
	\FloatBarrier

\subsubsection{Node set statistics} As a result of the actor natality and mortality described in the introduction, we can split the set of inventors $\mathcal{I}_t$ at each time step $t\in\mathcal{T}$ into new inventors with their first patent in $t$, $\mathcal{I}_t^+ = \{i \in \mathcal{I}_t; \sum_{u = t-s}^{t-1} \sum_{k \in \mathcal{K}_u} y_{u,ik} = 0\}$, and inventors that were already active prior to $t$, $\mathcal{I}_t^-= \{i \in \mathcal{I}_t; \sum_{u = t-s}^{t-1} \sum_{k \in \mathcal{K}_u} y_{u,ik} > 0\}$. We here use the term ``new inventors'' for actors in $\mathcal{I}_t^+$ and ``experienced inventors'' for those in $\mathcal{I}_t^-$. We can then define $\mathbf{y}_t^+ = (y_{t,ik})_{i \in \mathcal{I}_t^+, k \in \mathcal{K}_t}$ and $\mathbf{y}_t^- = (y_{t,ik})_{i \in \mathcal{I}_t^-, k \in \mathcal{K}_t}$ to be the sub-networks of $\mathbf{y}_{t}$ made up of new and experienced inventors, respectively.   

As it turns out, statistics on past behaviour such as 
\eqref{eq:stat_past_patents} are not meaningful for inventors from  $\mathcal{I}_t^+$, since no historical data is available for those inventors at time $t$. To account for this, we decompose the statistics $\mathbf{s}(\mathbf{y}_{t}, ..., \mathbf{y}_{t-s})$ into three types of terms, namely $\mathbf{s}^+(\mathbf{y}_t^+), \mathbf{s}^-(\mathbf{y}_t^-, ..., \mathbf{y}_{t-s})$, and $\mathbf{s}^\pm(\mathbf{y}_t)$, which are defined as statistics that only relate to either $\mathbf{y}_t^+$, $\mathbf{y}_t^-$ and past networks or the full set of inventors $\mathbf{y}_t$, respectively. Defining the corresponding coefficients ($\boldsymbol{\theta}^+,\boldsymbol{\theta}^-,\boldsymbol{\theta}^\pm$) and change statistics ($\boldsymbol{\Delta}_{t,ik}^+,\boldsymbol{\Delta}_{t,ik}^-,\boldsymbol{\Delta}_{t,ik}^\pm$) accordingly yields
\begin{align}
    \label{eq:change_stat_decomposability}
    \displaystyle
   \mathbb{P}_{\boldsymbol{\theta}}\left(Y_{t,ik} = 1| \mathbf{Y}_{t,ik}^C = \mathbf{y}_{t,ik}^C\right) =
   \begin{cases} \displaystyle
      \pi_{t,ik}^+(\mathbf{y}_t), &\text{if $i \in \mathcal{I}_t^+$ (new inventor)}\\
    \pi_{t,ik}^-(\mathbf{y}_t, ..., \mathbf{y}_{t-s}), &\text{if $i \in \mathcal{I}_t^-$ (experienced inventor)}
   \end{cases} 
\end{align}
where $\pi_{t,ik}^+(\mathbf{y}_t)$ and $\pi_{t,ik}^-(\mathbf{y}_t, ..., \mathbf{y}_{t-s})$ are given by 
\begin{align*}
    \pi_{t,ik}^+(\mathbf{y}_t) &= \frac{\exp\{(\boldsymbol{\theta}^+)^\top  \boldsymbol{\Delta}_{t,ik}^+(\mathbf{y}_t^+) + (\boldsymbol{\theta}^\pm)^\top  \boldsymbol{\Delta}_{t,ik}^\pm(\mathbf{y}_t)\}}{1 + \exp\{(\boldsymbol{\theta}^+)^\top  \boldsymbol{\Delta}_{t,ik}^+(\mathbf{y}_t^+)  + (\boldsymbol{\theta}^\pm)^\top  \boldsymbol{\Delta}_{t,ik}^\pm(\mathbf{y}_t)\}} \\
     \pi_{t,ik}^-(\mathbf{y}_t, ..., \mathbf{y}_{t-s}) &=  \frac{\exp\{(\boldsymbol{\theta}^-)^\top  \boldsymbol{\Delta}_{t,ik}^-(\mathbf{y}_t^-,..., \mathbf{y}_{t-s}) + (\boldsymbol{\theta}^\pm)^\top  \boldsymbol{\Delta}_{t,ik}^\pm(\mathbf{y}_t)\}}{1 + \exp\{(\boldsymbol{\theta}^-)^\top  \boldsymbol{\Delta}_{t,ik}^-(\mathbf{y}_t^-,..., \mathbf{y}_{t-s}) + (\boldsymbol{\theta}^\pm)^\top  \boldsymbol{\Delta}_{t,ik}^\pm(\mathbf{y}_t)\}}.
\end{align*}
As an example, for the common edge statistic $s_{\mathtt{edges}}(\mathbf{y}_t, ..., \mathbf{y}_{t-s})$, the aforementioned decomposition means we can define $s_{\mathtt{New}}(\mathbf{y}_t^+) = |\mathbf{y}_t^+|$ and $s_{\mathtt{Experienced}}(\mathbf{y}_t^-, ..., \mathbf{y}_{t-s}) = |\mathbf{y}_t^-|$, to allow for {new} and  {experienced} inventors to generally have a different propensity to be part of a patent. 
Note that the splitting of the node set as in \eqref{eq:change_stat_decomposability} does not assume any (in)dependence structure between $\mathbf{Y}_t^+$ and $\mathbf{Y}_t^-$, but rather serves as an aid to specify additional terms and interpret the coefficients at a finer level, as just exemplified for the edge statistic. 


\subsubsection{Adjustment for varying network size:} As argued in \citet{Krivitsky2011}, the task of comparing estimated coefficients of two models with identical specifications but different network sizes is non-trivial.  This behaviour is due to the fact that including the edge count statistic from the previous paragraph in a TERGM assumes density invariance as the network grows. This characteristic seldom holds for real-world networks as it implies a linearly growing mean degree of all involved actors. In the case of our longitudinal patent network, the number and composition of inventors and patents change from year to year, thus correcting for this is of practical importance to be able to compare coefficient estimates at different time points. To solve the issue, we follow the suggestion of \citet{Krivitsky2011} and incorporate the offset term $\frac{1}{|\mathcal{I}_t| + |\mathcal{K}_t|}$ to achieve asymptotically constant mean-degree scaling as the composition of inventors and patents change over time. 
	\FloatBarrier

\subsection{Estimation and inference}
\label{sec:estimation}
	\FloatBarrier

We now seek to estimate the parameter $\boldsymbol{\theta}$, by maximising the logarithmic likelihood constructed from \eqref{eq:tergm} for the transition between time points $t-1$ and $t$. To do so, we follow the Markov Chain Monte Carlo Maximum Likelihood Estimation procedure introduced by \citet{GeyerThompson:1992} and adapted to ERGMs by  \citet{hunter2006inference}. In our application, we repeat this for each available time step $t = 1, ..., T$.

First, note that subtracting any constant from the logarithmic  likelihood constructed from \eqref{eq:tergm} does not change its maximum. We can therefore subtract the logarithmic likelihood evaluated at an arbitrary value of the parameter $\boldsymbol{\theta}$, i.e.\ $\boldsymbol{\theta}_{0}$, which yields the equivalent objective function
\begin{align}\label{eq:loglik_mcmcmle}
l(\boldsymbol{\theta}) - l(\boldsymbol{\theta}_{0}) = (\boldsymbol{\theta} - \boldsymbol{\theta}_{0})^\top\boldsymbol{s}(\mathbf{y}_{t},\ldots,\mathbf{y}_{t-s}) - \log\left(\mathbb{E}_{\boldsymbol{\theta}_{0}}\left(\exp\{ (\boldsymbol{\theta}-\boldsymbol{\theta}_{0})^\top\boldsymbol{s}(\mathbf{Y}_{t},\ldots,\mathbf{y}_{t-s})  \}\right)\right),
\end{align}
where $\mathbb{E}_{\boldsymbol{\theta}}(f(\mathbf{X}))$ is the expected value of random variable $\mathbf{X}$ characterised by parameter $\boldsymbol{\theta}$ and transformed through the arbitrary function $f(\cdot)$. As described in \citet{hunter2006inference}, one can evaluate this objective function by approximating the expected value by generating random networks $\mathbf{Y}^{(1)},\mathbf{Y}^{(2)},\ldots,\mathbf{Y}^{(M)}$ from \eqref{eq:tergm} under $\boldsymbol{\theta}_0$. In particular, we approximate the expected value in \eqref{eq:loglik_mcmcmle} through a Monte Carlo quadrature:
\begin{align}\label{eq:loglik_mcmcmle_approx}
\mathbb{E}_{\boldsymbol{\theta}_{0}}\left(\exp\{ (\boldsymbol{\theta}-\boldsymbol{\theta}_{0})^{\top}\boldsymbol{s}(\mathbf{Y}_{t},\ldots,\mathbf{y}_{t-s})  \}\right) \approx \frac{1}{M} \sum_{m=1}^{M}\exp\left\{ (\boldsymbol{\theta} - \boldsymbol{\theta}_{0})^{\top}\boldsymbol{s}(\mathbf{y}^{(m)},\mathbf{y}_{t-1}, ..., \mathbf{y}_{t-s}) \right\}
\end{align}
For sufficiently large $M$ the convergence of this expectation is guaranteed, and we can plug \eqref{eq:loglik_mcmcmle_approx} into \eqref{eq:loglik_mcmcmle} and apply Newton-Raphson type of methods to maximise it with respect to $\boldsymbol{\theta}$. Sampling from a probability distribution with intractable normalisation constant, such as \eqref{eq:tergm}, is achieved by a Metropolis-Hastings algorithm. In particular, we first sample an edge, defined as the tuple $(i,k)$, at random, and consecutively toggle the corresponding entry of $\mathbf{Y}_t$ from 0 to 1 with probability equal to \eqref{eq:change_stat} (for more details see \citealp{hunter2013}). Due to the large size of the patent networks, we propose 15.000 of such changes and then stop the Markov chain. This procedure is hence equivalent to contrastive divergence \citep{Krivitsky2017}. 

Inference  on the estimates is drawn based on the Fisher matrix $\mathbf{I}(\boldsymbol{\theta})$, which equals the variance of the sufficient statistics for exponential family distributions \citep{Wassennan2004}. Thus, we can approximate the Fisher matrix through
\begin{align*}\label{eq:loglik_mcmcmle_approx}
\widehat{\mathbf{I}}(\boldsymbol{\theta}) = \text{Var}_{\theta} (\boldsymbol{s}(\mathbf{Y}_{t},\ldots,\mathbf{y}_{t-s})) \approx \frac{1}{M} \sum_{m=1}^{M} &
\left(\boldsymbol{s}\left(\mathbf{y}^{(m)}, \mathbf{y}_{t-1}, ..., \mathbf{y}_{t-s}\right) - \bar{\boldsymbol{s}}\left(\mathbf{y}^{(1)}, ..., \mathbf{y}^{(M)}\right)\right)\\ &\left(\boldsymbol{s}\left(\mathbf{y}^{(m)}, \mathbf{y}_{t-1}, ..., \mathbf{y}_{t-s}\right) - \bar{\boldsymbol{s}}\left(\mathbf{y}^{(1)}, ..., \mathbf{y}^{(M)}\right)\right)^\top 
\end{align*}
where $\displaystyle\bar{\boldsymbol{s}}\left(\mathbf{y}^{(1)}, ..., \mathbf{y}^{(M)}\right) = \dfrac{1}{M} \sum_{m = 1}^M \boldsymbol{s}\left(\mathbf{y}^{(m)}, \mathbf{y}_{t-1}, ..., \mathbf{y}_{t-s}\right)$ is the vector of average of the sufficient statistics from the simulated networks $\mathbf{y}^{(1)},\ldots,\mathbf{y}^{(M)}$, which are, in turn, drawn from the fitted model with parameter $\boldsymbol{\theta}$ set to its maximum likelihood estimate. 

	\FloatBarrier

\section{Application}
\label{sec:application}

We can now present the results of the application of our model to the patent data introduced in Section \ref{sec:patent_data}. For each statistic included in the model, we first explain its meaning and subsequently interpret the corresponding estimated coefficient. Further details on the specification of each sufficient statistic can be found in Appendix A. MCMC diagnostics, and goodness-of-fit assessments as proposed by \citet{Hunter2008b} are provided in the Supplementary Material. 


\subsection{Network effects}
\label{sec:specification}




\begin{figure}[t!]\centering
	\FloatBarrier
	\includegraphics[trim={0cm 0cm 0cm 0cm},clip,width=0.95\textwidth]{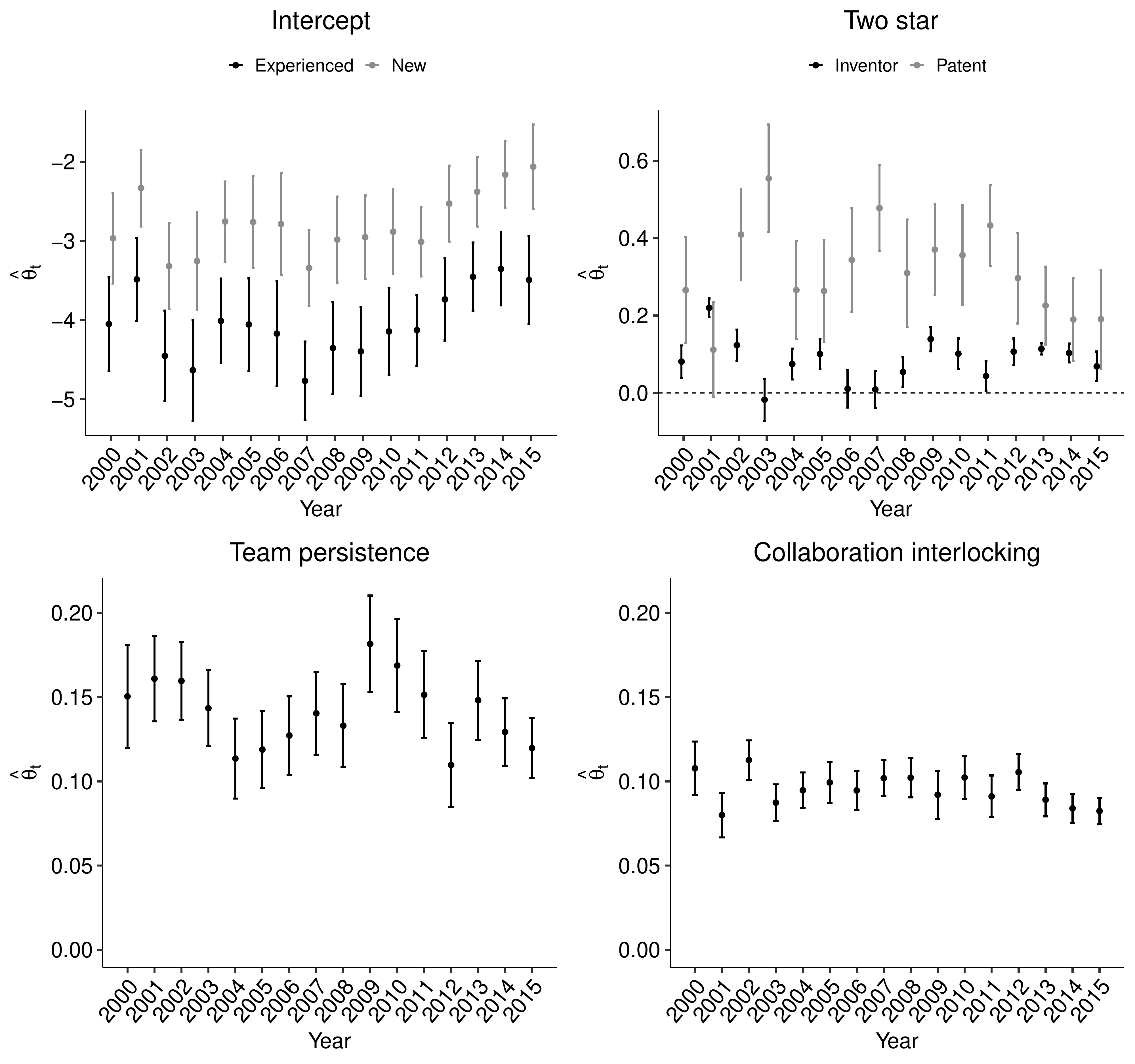}
			\caption{Estimated time-varying coefficients regarding the propensity to invent, two-star statistics, team persistence and collaboration interlocking.}
	\label{fig:res1}
\end{figure}

\textbf{Propensity to invent:} To account for the changing activity levels over time, we incorporate a statistic that counts how many edges are in the network. Following Section \ref{sec:sufficient}, we split this term into separate statistics for experienced and new inventors. Heuristically, one can interpret the corresponding coefficients as the general propensity to form ties, i.e.\ participate in a patent, for the two inventor sets, respectively. 
The plot of the corresponding estimates is shown in the upper left panel of Figure \ref{fig:res1}. It exhibits a different level of activity for new and experienced inventors. We expect this by design, as new inventors enter the network precisely because they are active at time $t$, while experienced ones might only have been active in the past. Overall, we observe a steady increase in activity in the network over time from 2008 onward for both sets of inventors.



\begin{figure}
    \centering\captionsetup[sub]{font=large}
    \begin{subfigure}[c]{0.4\textwidth}
    \includegraphics[width=\textwidth, page = 2]{fig/statistics.pdf}
    \subcaption{Team persistence 
    }
    \label{fig:rep}
    \end{subfigure}
    \begin{subfigure}[c]{0.4\textwidth}
    \includegraphics[width =\textwidth, page = 1]{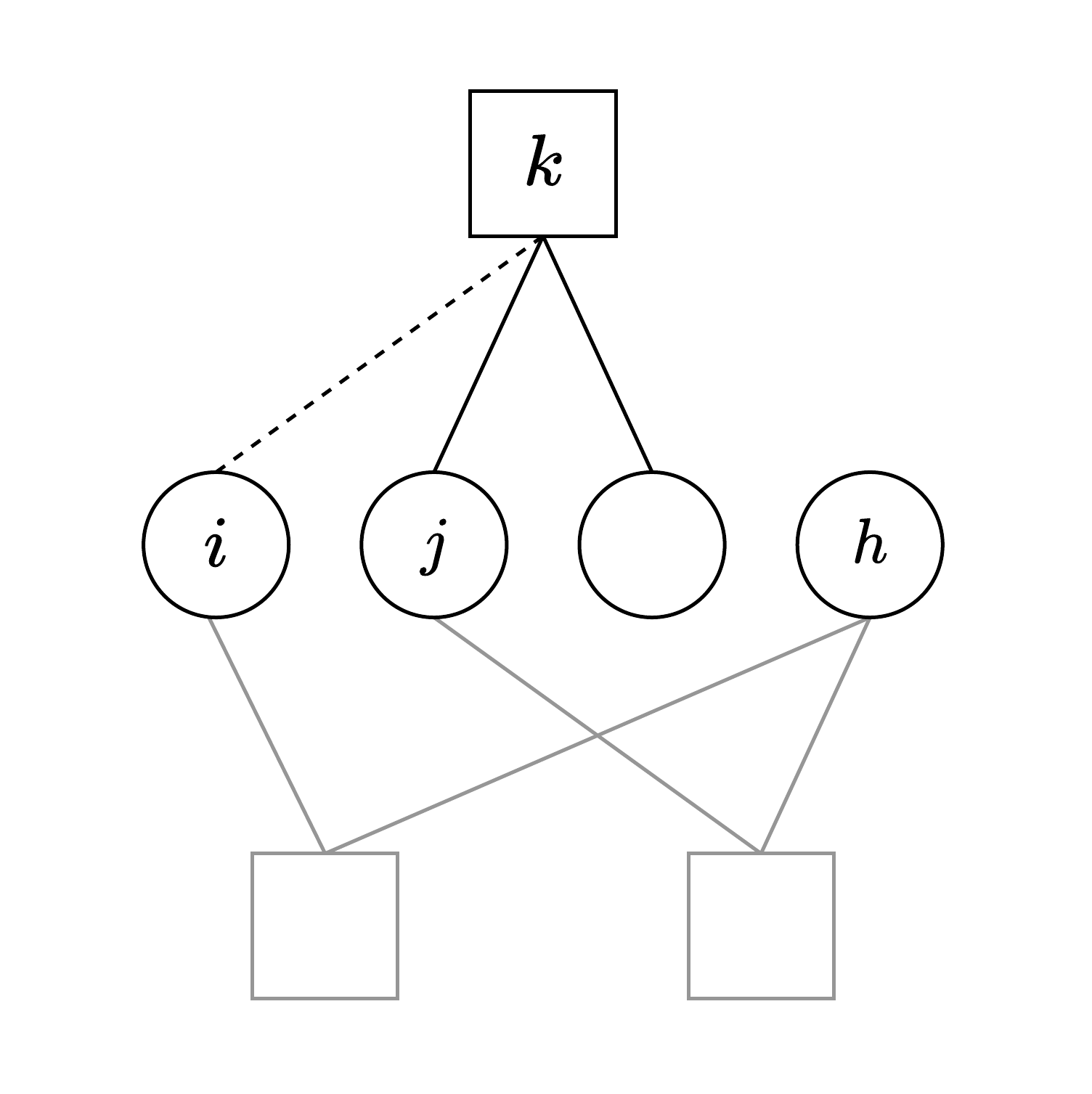}
    \subcaption{{Collaboration interlocking
    }}
    \label{fig:common}
    \end{subfigure}
    \caption{Illustration of the change statistics related to assortative network statistics for team persistence (a) and collaboration interlocking (b). Circles represent inventors, and squares are patents. The dashed line indicates a possible edge at time point $t$, while black lines represent edges given at time point $t$. Grey lines, on the other hand, display past connections, and grey squares stand for past patents.}
    \label{fig:rep_common}
\end{figure}

\textbf{Two-star statistics:} Two-star statistics relate to the concept of centrality \citep{Wasserman1994}. For bipartite networks, they can be defined with respect to each of the two modes (inventors and patents, respectively). For inventors the statistic is given in Appendix A and expresses whether inventor $i$ is more or less likely to invent an additional patent in year $t$, given that he/she is (co-)owner of at least another patent in that year. For patents, the statistic relates to the number of inventors per patent and is given in \eqref{eq:2star}. 
 The top right panel of Figure \ref{fig:rep_common} depicts the two corresponding estimates. For inventors, the estimates take for most time points small positive values without much temporal variation. This indicates  a slight tendency towards centralisation for inventors, i.e.\ inventors aiming to submit multiple patents per year. For patents the corresponding two-star estimates are larger, i.e.\ patents tend to be owned by multiple inventors. The two star effect slowly decreases since 2011, meaning that the number of owners per patent is getting smaller. 
 The variance for the estimated two-star patent effect is generally larger than the estimate of the corresponding two-star inventor effect, which  stems from the fact that there are fewer patents than inventors in a single year.


\textbf{Team persistence:}
Most patented inventions are the result of team work \citep{GIURI20071107}, which leads to the build-up of valuable team-specific capital \citep{Jaraveletal2018}. We therefore expect past collaboration to positively affect the propensity for two inventors to collaborate again. We include the team persistence statistic based on the pairwise statistics of inventors proposed in \ref{sec:sufficient} in the model to account for this effect. The statistic, which could also be defined as ``repetition statistic'', is visually represented in Figure \ref{fig:rep_common} (a), and rests on the definition of matrix $\mathbf{x}_{t}^{\mathtt{P}}$, whose $(i,j)$th entry is 1 if inventor $i$ and $j$ have already co-invented a patent in the previous five years and 0 otherwise. The bottom left panel of Figure \ref{fig:res1} depicts the corresponding coefficient estimate, which is significantly positive over time. This finding corroborates our anticipations that, controlling for the other factors, two inventors are more likely to jointly produce a patent if they already worked on an invention together in the past. Hence teams of inventors play an important role in patent creation.

\textbf{Collaboration interlocking:}
In addition to investigating the persistence of collaborations, it is also of interest to understand how having had a common partner in the past influences the tendency to develop a patent together in the present. We account for this by including the collaboration interlocking statistic in our model. By common partners we are referring to actors such as inventor $h$ for inventors $i$ and $j$ in Figure \ref{fig:rep_common} (b). We define the statistic again by pairwise statistics of inventors through the matrix $\mathbf{x}_{t}^{\mathtt{CI}}$, encoding in the $(i,j)$th entry the binary information if inventors $i$ and $j$ have at least one common partner or not. The related coefficient estimates are shown in the bottom right panel of Figure \ref{fig:res1}, where we notice that the estimate attains significantly positive values throughout the observational period. This result suggests that if two inventors $i$ and $j$ both had a patent with the same inventor $h$ in the past, they are generally more likely to co-invent in the future. The finding holds controlling for all other features in our model (including the previously described team persistence statistic), and can be viewed as akin to triadic closure in unimodal networks, i.e.\ ``a collaborator of my collaborator is more likely to become my collaborator''. The result thus supports the idea that the creation of inventor teams is often promoted via common colleagues and that informal knowledge flows are key to the invention process (see \citealp{giuri2013distance} and references cited therein).

\subsection{Effects of inventor-specific covariates}

\begin{figure}[t!]\centering
	\FloatBarrier
	\includegraphics[trim={0cm 0cm 0cm 0cm},clip,width=0.95\textwidth]{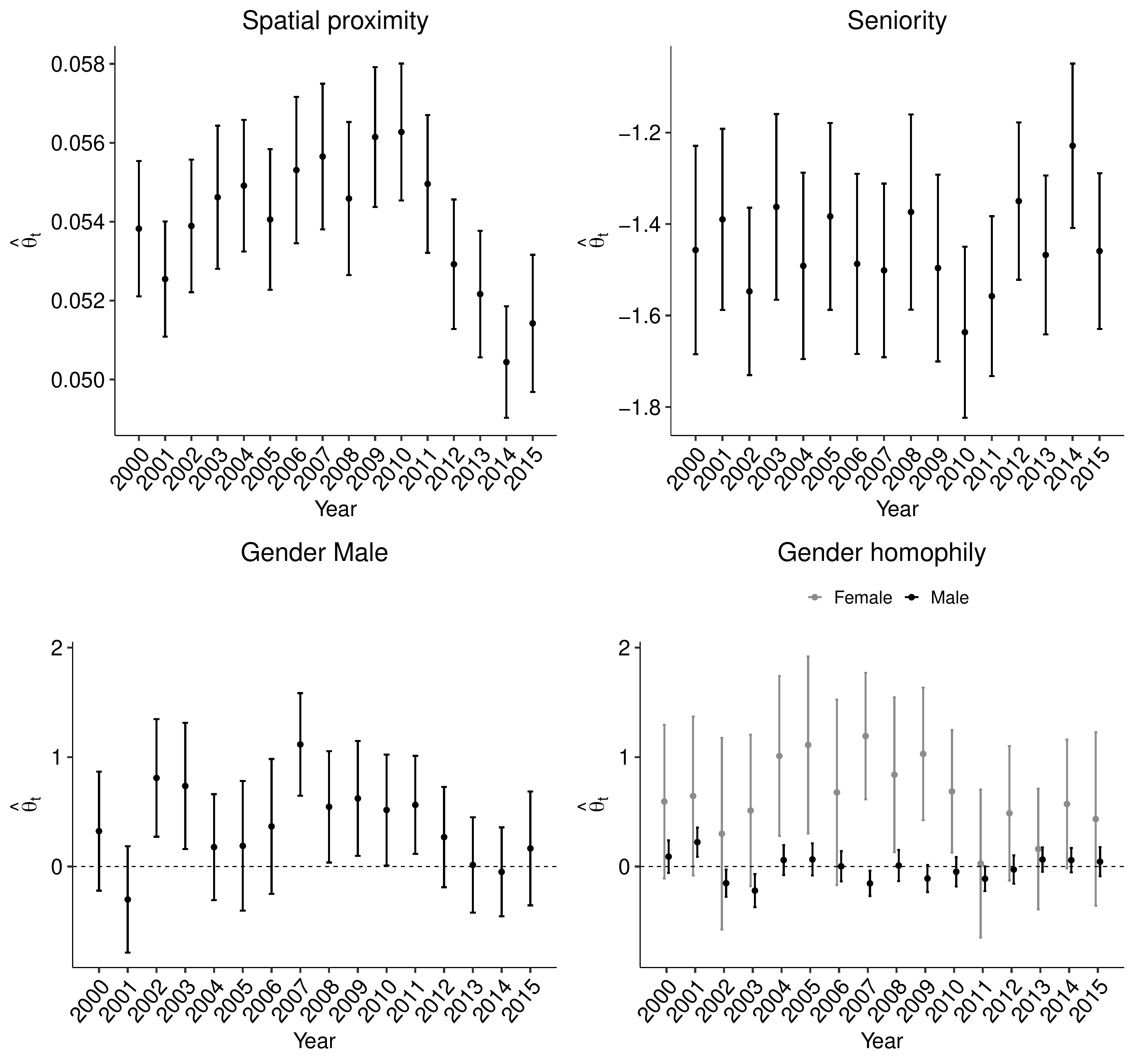}
	\caption{Estimated time-varying coefficients regarding the spatial proximity, seniority, and gender of inventors.}
	\label{fig:res2}
\end{figure}

\textbf{Spatial proximity:}
Many patents are created in a workplace environment \citep{GIURI20071107}. For this reason, we would expect inventors that live close to each other to be more likely to invent together. Moreover, there is empirical evidence that collaboration is more likely between inventors that live close to one another even if they do not share the same employer (e.g., \citealp{CRESCENZI2016177}). For these reasons, we include a spatial proximity statistic in our model, where we define spatial proximity as living within a range of 50 km. We encode this proximity information in a binary matrix $\mathbf{x}^{\mathtt{SP}}$ and incorporate it in the model as a pairwise statistic of inventors. The top left panel of Figure \ref{fig:res2} depicts the estimated coefficients for the statistic. The positive values attained over time confirm that inventors living near each other have a higher chance to collaborate. We can also see that the effect goes down over time from 2010 onward; this makes sense in an increasingly interconnected society, where more and more connections are formed through the web in addition to physical ones.

\textbf{Seniority:}
The top right panel in Figure \ref{fig:res2} depicts the effect of the number of previously owned patents by each inventor in the past five years. The corresponding statistic can be viewed as a measure of inventor seniority, where inventors with more patents in the past are considered to be senior. Note that this statistic is only computed for the set of inventors which were previously active in the network (experienced inventors), as for new inventors it would trivially be a structural zero. The negative coefficient estimate here suggests that, conditional on all other statistics included in the model, senior inventors have a lower propensity to create new patents. Prior research has shown that career dynamics of inventors are complex as economic opportunities, productivity and personal preferences interact (see, e.g., \citealp{allen1992age} and \citealp{bell2019tax}). But our results would be consistent with earlier results that with greater seniority, inventors take over  managerial responsibilities within the same firm, or that high visibility of their invention output also leads them to move to new employers and tasks, thus lowering (or halting) their invention output.


\textbf{Gender and gender homophily:} Another variable of interest in the realm of innovation research is gender. Many researchers have expressed concerns about the extremely low representation of women among inventors (typically far less than 10 $\%$) and possible wage discrimination, see, e.g., \citet{hoisl2017sa} and \citet{jensen2018gender}. These studies made  gender an important topic in innovation economics. We incorporate gender in our model in two ways, i.e. as a main effect and as homophily effect introduced in \eqref{eq:2star_homo}. 
The two plots at the bottom of Figure \ref{fig:res2} show the effects of gender on the propensity to create patents (left) and on homophily, i.e. the tendency of inventing together with people of the same gender (right). Note that both effects need to be interpreted keeping in mind that the vast majority of the actors in the network are male (96\%). From the plot on the bottom left, we can see how, while male inventors seem to be slightly more active, all in all male and female inventors did not show significant differences in their propensity to invent. Note that this holds given the inclusion of those inventors in the network, i.e. given that they were already inventors. The gender homophily plot shows different results; here we see that, while male inventors seem to have the same likelihood to form patents with both genders, female inventors tend to have more collaborations with other females than with males. While the effect is quite sizeable in absolute value,  the uncertainty here is considerable given the small number of female actors in the network. Still, we can see this as weak evidence for a gender homophily effect for female inventors. These results are consistent with earlier findings by \citet{WHITTINGTON2018511} who studies the role of gender in life science inventor teams. 

\section{Discussion}
\label{sec:discussion}
This paper analyses a massive bipartite network consisting of all inventors and collaborative patents submitted between 1995 and 2015 in electrical engineering. To account for the sheer size of the complete network and the structural zeros in the related bipartite adjacency matrix, we suggested a temporal decomposition of the data into multiple smaller networks. Guided by substantive questions posed by innovation research, we then proposed a set of novel bipartite network statistics focused on gender issues, team persistence, collaboration interlocking and spatial proximity. 

Time-varying actor sets due to actor mortality and natality are often observed in networks beyond the realm of patent data. For instance, scientific collaboration behaves similarly, as many PhD students do not pursue an academic career and hence have a short lifespan in the scientific collaboration network. At the same time, new PhD students continuously enter the scientific world. Therefore, we argue that the proposed temporal decomposition and the novel class of network terms can be employed in other settings. 

In addition to the methodological contributions, our study offers several empirical findings. In particular, we show how spatial proximity, team work and interlocking of collaborations each have a positive impact on the output of inventors. We also demonstrate how inventors' characteristics, such as gender and seniority, play a significant role in the process. 

All in all, our application to inventor teams presents an alternative to classical forms of analysis of patenting and inventorship networks. While prior studies are almost exclusively focused on analysing the underlying mechanisms one at a time, we model them simultaneously in the framework of bipartite networks. We argue that this can provide an effective alternative to classical forms of regression-based analysis of this important phenomenon.




\section*{Funding}
The work has been partially supported by the German Federal Ministry of Education and Research (BMBF) under Grant No. 01IS18036A. The authors of this work take full responsibility for its content.

	\bibliographystyle{chicago}
	
	\bibliography{library}
\newpage
\section*{A Sufficient statistics}
In the following, we detail the mathematical definitions of all sufficient statistics incorporated in our model. 

\begin{figure}
    \centering\captionsetup[sub]{font=large}
    \begin{subfigure}[c]{0.3\textwidth}
    \includegraphics[width=\textwidth, page = 7]{fig/graphs_uni.pdf}
    \subcaption{Experienced inventors}
    \label{fig:rep}
    \end{subfigure}
    \begin{subfigure}[c]{0.3\textwidth}
   \includegraphics[width=\textwidth, page = 8]{fig/graphs_uni.pdf}
    \subcaption{New inventors}
    \label{fig:common}
    \end{subfigure}
     \begin{subfigure}[c]{0.3\textwidth}
    \includegraphics[width=\textwidth, page = 1]{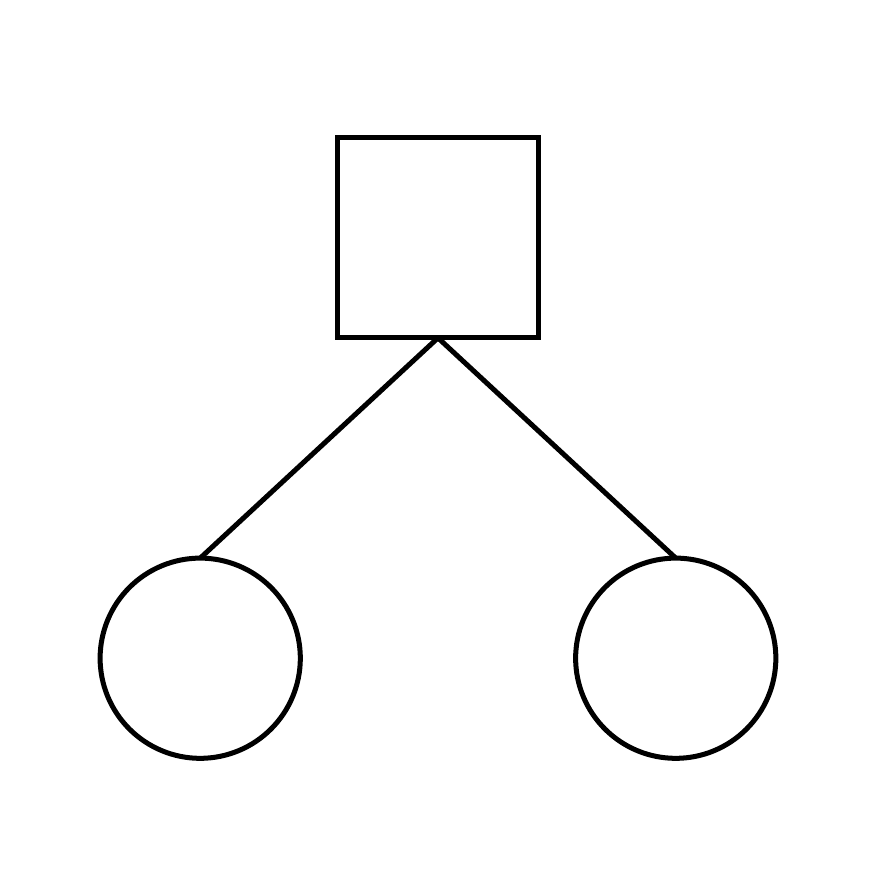}
    \subcaption{Patent two-stars}
    \label{fig:rep}
    \end{subfigure}
    \begin{subfigure}[c]{0.3\textwidth}
   \includegraphics[width=\textwidth, page = 2]{fig/graphs_uni.pdf}
    \subcaption{Inventor two-stars}
    \label{fig:common}
    \end{subfigure}
    \begin{subfigure}[c]{0.3\textwidth}
    \includegraphics[width=\textwidth, page = 5]{fig/graphs_uni.pdf}
    \subcaption{Seniority}
    \label{fig:rep}
    \end{subfigure}     
    \begin{subfigure}[c]{0.3\textwidth}
    \includegraphics[width=\textwidth, page = 6]{fig/graphs_uni.pdf}
    \subcaption{Male inventors}
    \label{fig:rep}
    \end{subfigure}
    \begin{subfigure}[c]{0.34\textwidth}
    \includegraphics[width=\textwidth, page = 1]{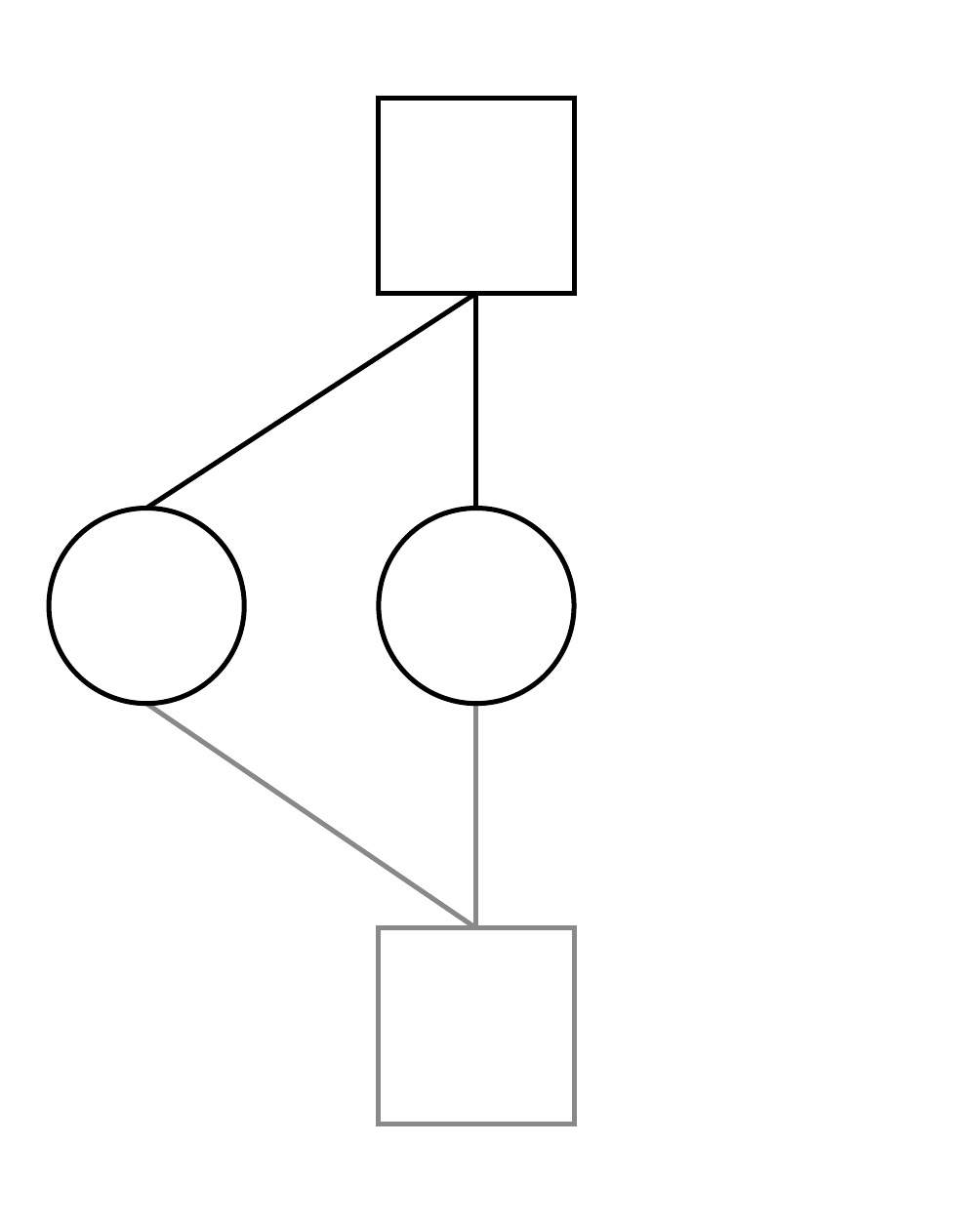}
    \subcaption{Team persistence}
    \label{fig:rep}
    \end{subfigure}
    \begin{subfigure}[c]{0.34\textwidth}
    \includegraphics[width=\textwidth, page = 2]{fig/fig_bi.pdf}
    \subcaption{Collaboration interlocking}
    \label{fig:common}
    \end{subfigure}
    \begin{subfigure}[c]{0.3\textwidth}
       \vspace{-0.19cm}
    \includegraphics[width=\textwidth, page = 4]{fig/graphs_uni.pdf}
    \vspace{1.7cm}
    \subcaption{Homophily of females}
    \label{fig:rep}
    \end{subfigure}
    \caption{Network configurations for the general edge and two-star terms. Circles are inventors and squares patents and black lines are observed edges in the network at time point $t$, while grey lines are edges in the past.}
    \label{fig:fig_main}
\end{figure}

\textbf{Propensity to invent:} As already stated in Section \ref{sec:tergm}, the standard term to incorporate in any ERGM specification is an edge statistic that counts how many edges are realised in the network. In accordance with Section \ref{sec:sufficient}, we split this term into the statistics $s_{\mathtt{New}}(\mathbf{y}_t^+) = |\mathbf{y}_t^+| = \sum_{i \in \mathcal{I}^+_t} \sum_{k \in \mathcal{K}_t} y_{t,ik}$ and $s_{\mathtt{Experienced}}(\mathbf{y}_t^-, ..., \mathbf{y}_{t-s})= |\mathbf{y}_t^-| =\sum_{i \in \mathcal{I}^-_t} \sum_{k \in \mathcal{K}_t} y_{t,ik}$. Figures \ref{fig:fig_main}(a) and \ref{fig:fig_main}(b) visualise the corresponding two network configurations. 


\textbf{Two-star statistics:} Two-star statistics can be stated regarding either set of actors in the case of bipartite networks. The definition of the two-star statistic for the patents  is shown in Figure \ref{fig:fig_main}(c) and given by
\begin{align*}
    s_{\mathtt{twostar.patent}}(\mathbf{y}_t) = \frac{1}{2} 
    \sum_{k \in \mathcal{K}_t} \sum_{i \in \mathcal{I}_t} 
    y_{t,ik} \left(\sum_{j\neq i} y_{t,jk}\right),
\end{align*}
while the version for the inventors is visualised in Figure \ref{fig:fig_main}(d) and defined as: 
\begin{align*}
    s_{\mathtt{twostar.inventor}}(\mathbf{y}_t) = \frac{1}{2} 
    \sum_{i \in \mathcal{I}_t} \sum_{k \in \mathcal{K}_t} 
    y_{t,ik} \left(\sum_{l\neq k} y_{t,il}\right).
\end{align*}

\textbf{Pairwise statistics of inventors:} We include three versions of pairwise statistics of inventors introduced in Section \ref{sec:sufficient}. The statistics are given by 
\begin{align*}
    s_{\mathtt{assort.x}}(\mathbf{y}_t, ..., , \mathbf{y}_{t-s}) =\frac{1}{2} \sum_{k \in \mathcal{K}_t} \sum_{i \in \mathcal{I}_t} y_{t,ik} \left(100\times \frac{ \sum_{j\neq i} y_{t,jk}x_{t,ij}}{    \sum_{j\neq i} y_{t,jk}}\right).
\end{align*}
Note that, in general, the matrix $\mathbf{x}$ can be an arbitrary function of the past networks and nodal or dyadic exogenous information. Its definition differs between the three statistics of pairwise statistics of inventors:
\begin{enumerate}
    \item Team persistence: For $i,j \in \mathcal{I}_t$ and $i\neq j$ the entries of $\mathbf{x}_t^{P}$ are given by
    \begin{align*} \displaystyle
        x_{t,ij}^P = \begin{cases}1,& \text{if } \sum_{u = t-s}^{t-1} \sum_{k\in \mathcal{K}_u} y_{u,ik}y_{u,jk} > 0 \\
           0,&\text{else}
           \end{cases}
    \end{align*}
    and a graphical illustration of the statistic is provided in Figure \ref{fig:fig_main}(g). One can comprehend this statistic as a particular type of the four-cycle statistic \citep{Wang2013} where one half already occurred in the past, and the other half might occur in the present.
    \item Collaboration interlocking: For $i,j \in \mathcal{I}_t$ and $i\neq j$ the entries of $\mathbf{x}_t^{CI}$ are defined by
    \begin{align*}
        x_{t,ij}^{CI} = \begin{cases}1,& \text{if } \sum_{u = t-s}^{t-1} \sum_{h\in \mathcal{I}_t}\sum_{k,l\in \mathcal{K}_u} y_{u,ik} y_{u,hk} y_{u,jl} y_{u,hl} > 0 \\
           0,&\text{else}
           \end{cases}
    \end{align*}
    and a graphical illustration of the statistic is provided in Figure \ref{fig:fig_main}(h). Coming back to the representation as cycle-statistics, this term is a six-cycle statistic in which four of the six edges happened in the time frame from $t-5$ to $t-1$ and two in year $t$.
    \item Spatial proximity: For $i,j \in \mathcal{I}_t$ and $i\neq j$ the entries of $\mathbf{x}_t^{SP}$ are defined as
    \begin{align*}
        x_{t,ij}^{SP} = \begin{cases}1,& \text{if } \text{dist}(x_{\mathtt{coord},i},x_{\mathtt{coord},j}) > 50\text{km} \\
           0,&\text{else}
           \end{cases}
    \end{align*}
    where $x_{coord,i}$ and $x_{coord,j}$ define the longitude and latitude of inventors $i$ and $j$, respectively, and the function  $\text{dist}(x_{coord,i},x_{coord,j})$ computes the distance in kilometers between them via the haversine formula.  
\end{enumerate}

\textbf{Seniority:} The respective binary indicator is based on the $\mathtt{pastpatent}$ statistic given in \eqref{eq:stat_past_patents}, but in this case we define it on the inventor level:
\begin{align*}
    s_{\mathtt{seniority},i}(\mathbf{y}_{t}, ..., \mathbf{y}_{t-s}) = \sum_{u = t-s}^{t-1} \sum_{k \in \mathcal{K}_u} y_{u,ik}
\end{align*}
We binarise this inventor-specific covariate by first computing the median of $s_{\mathtt{seniority},i}(\mathbf{y}_{t}, ..., \mathbf{y}_{t-s})$ over all inventors and then using this value to split the inventors into two groups (i.e. {seniors} and {juniors}). The resulting categorical covariate relates to the number of patents in the past, and is represented in Figure \ref{fig:fig_main}(e).

\textbf{Gender and gender homophily:} The main effect of gender is depicted in Figure \ref{fig:fig_main}(f) and defined by: 
\begin{align*}
    s_{\mathtt{gender}}(\mathbf{y}_t) = \sum_{i \in \mathcal{I}_t} \sum_{k \in \mathcal{K}_t} y_{t,ik}\mathbbm{1}(x_{\mathtt{gender},i} = \text{``male''}),
\end{align*}
where $x_{\mathtt{gender},i} \in \{\text{``male''},\text{``female''}\}$ indicates the gender of inventor $i$. The homophily effect, on the other hand, is for males defined by: 
\begin{align}
    s_{\mathtt{homophily.male}}(\mathbf{y}_t) = \frac{1}{2} \sum_{k \in \mathcal{K}_t} \sum_{i \in \mathcal{I}_t} y_{t,ik} \left(\sum_{j\neq i} y_{t,jk} \mathbbm{1}(x_{\mathtt{gender},i} =\text{``male''})\mathbbm{1}(x_{\mathtt{gender},j} =\text{``male''})\right).
\end{align} 
and for females the formula reads : 
\begin{align}
    s_{\mathtt{homophily.female}}(\mathbf{y}_t) = \frac{1}{2} \sum_{k \in \mathcal{K}_t} \sum_{i \in \mathcal{I}_t} y_{t,ik} \left(\sum_{j\neq i} y_{t,jk} \mathbbm{1}(x_{\mathtt{gender},i} =\text{``female''})\mathbbm{1}(x_{\mathtt{gender},j} =\text{``female''})\right).
\end{align} 
 Figure \ref{fig:fig_main}(i) visualises the homophily statistic for females. The equivalent statistic for males can be defined in the same manner. 

\newpage
\begin{center}
	\textbf{Supplementary material: Modelling the large and dynamically growing bipartite network of German patents and inventors} \\ 
	Cornelius Fritz$^1$\footnote[1]{Corresponding Author:  \href{mailto:cornelius.fritz@stat.uni-muenchen.de}{cornelius.fritz@stat.uni-muenchen.de}}, Giacomo De Nicola$^1$, Sevag Kevork$^1$,\\ Dietmar Harhoff$^2$, Göran Kauermann$^1$\hspace{.2cm}\\
	\vspace{0.3cm}
	$^1$Department of Statistics, LMU Munich, Germany \\
	$^2$Max Planck Institute for Innovation and Competition, Munich, Germany\hspace{.2cm}
\end{center}

\setcounter{section}{0}
\setcounter{page}{1}

\section{Descriptive statistics of the patent networks}

\begin{figure}[t]
    \centering
    \includegraphics[width=0.45\textwidth]{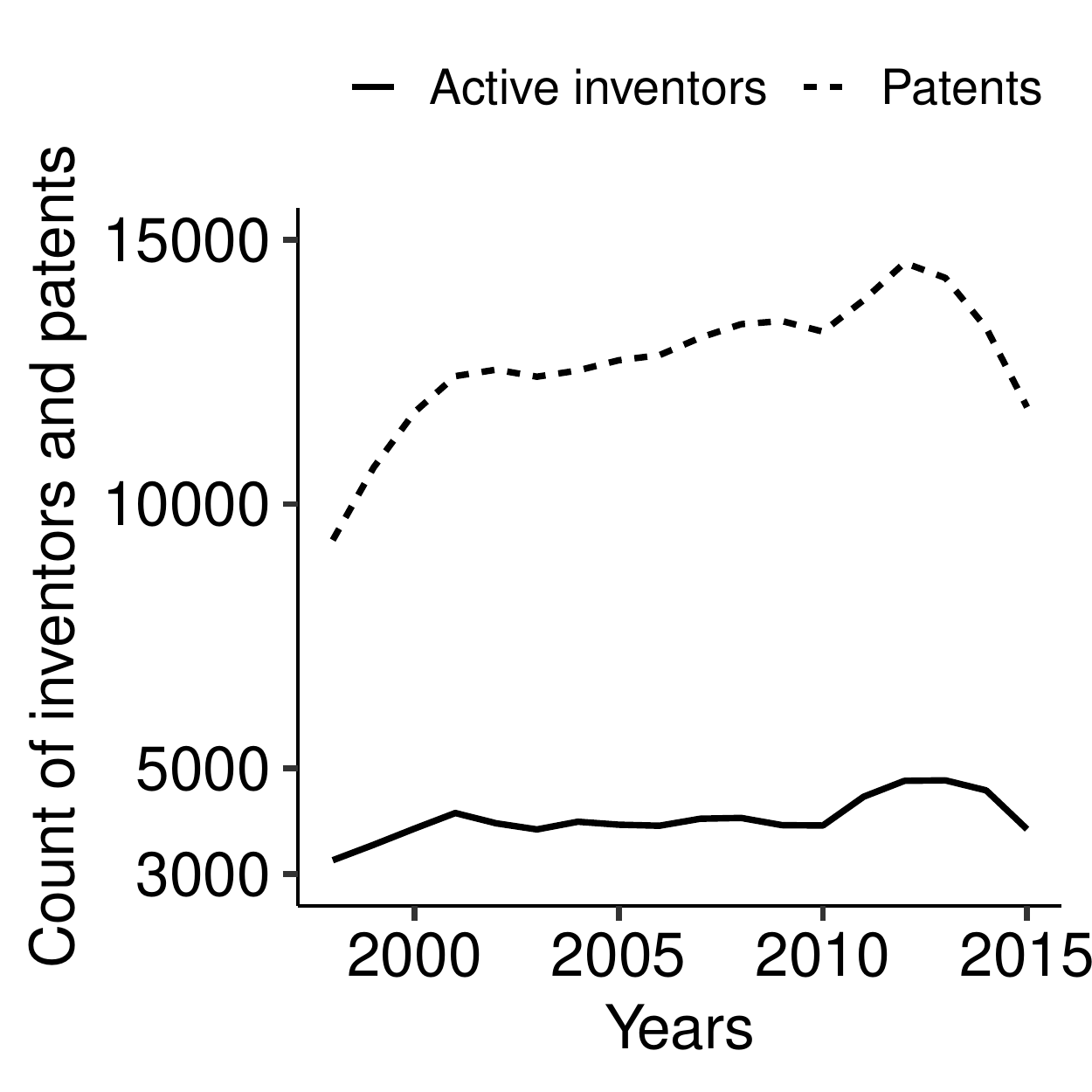}
    \includegraphics[width=0.45\textwidth]{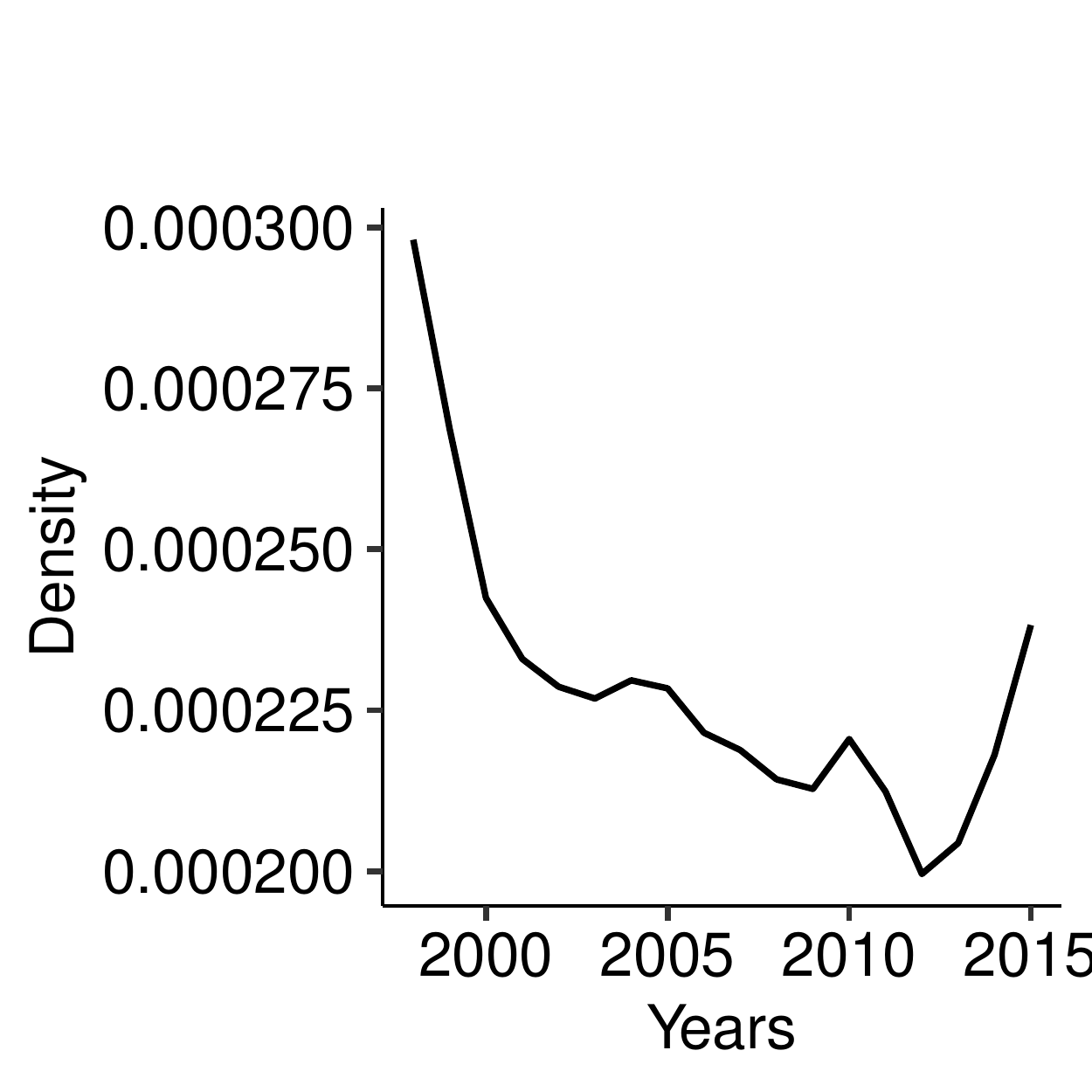}
    \includegraphics[width=0.45\textwidth]{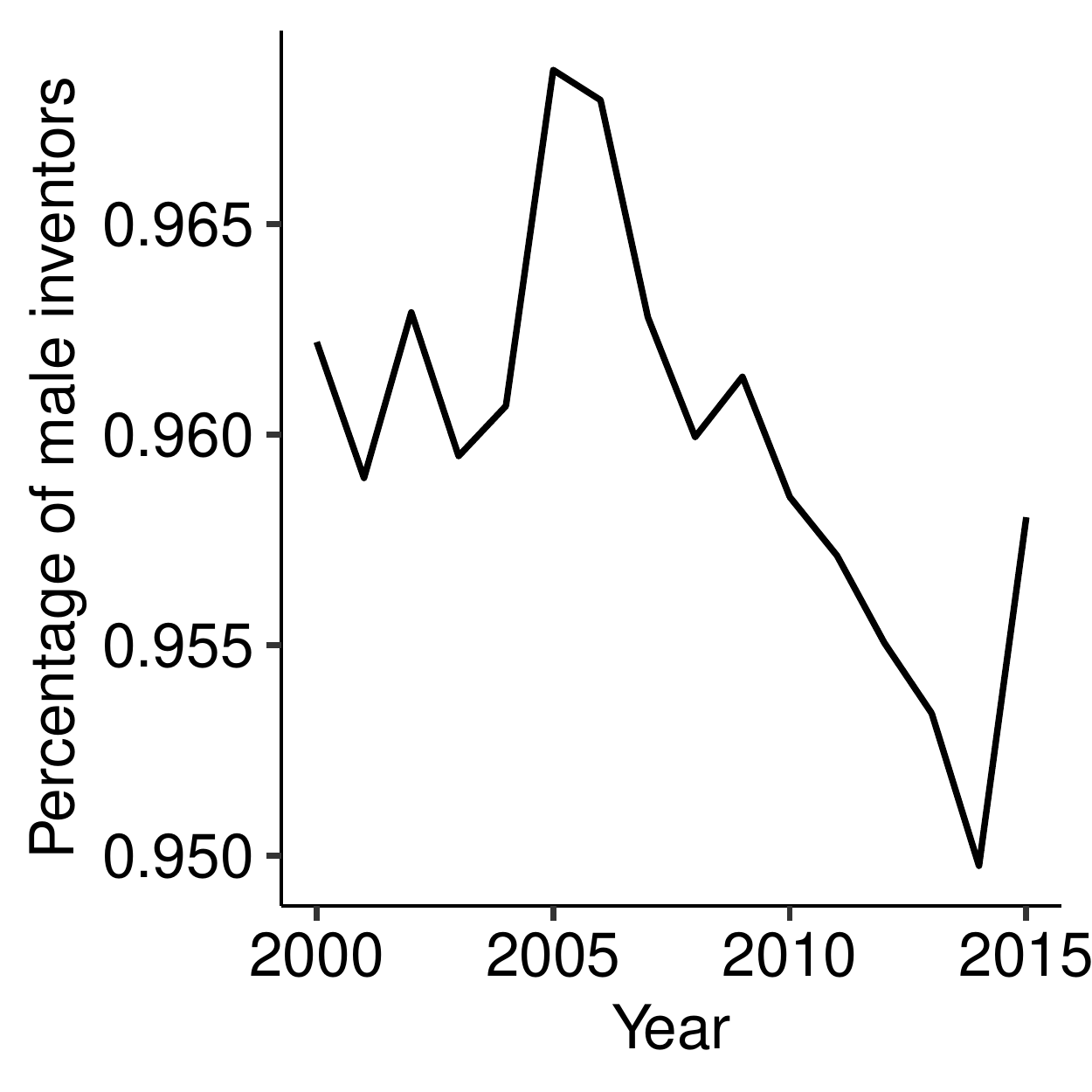}
    \caption{(a) Number of inventors and patents  over time. (b) Density of the patent network. (c) Percentage of male inventors over time.}
    \label{fig:stats}
\end{figure}

The top left panel of Figure \ref{fig:stats} depicts the absolute number of active inventors and patents in each year. In general, the number of patents per year increase over time, although we observe a drop in recent years. Moreover, there are fewer active inventors than patents in the networks, and the number of inventors is relatively constant over time. On the right side of  Figure \ref{fig:stats}, we plot the density over time. The density declines over time, although we see a boost in the last three years. 

\section{Implementation}

To carry out the analysis of the presented manuscript, we heavily rely on the $\mathtt{statnet}$ suite of packages. In particular, we implement wrapper functions to use the $\mathtt{ergm}$ package for the estimation  \citep{Hunter2008,Krivitsky2021} and the package template $\mathtt{ergm.userterms}$\citep{hunter2013} to employ the novel pairwise statistics of inventors. To enable the use of those statistics in other applications, we make the software package $\mathtt{patent.ergm}$ available for the software package $\mathtt{R}$ \citep{R_update}. Moreover, we can provide the complete replication code, including the goodness-of-fit analysis and MCMC diagnostics.

\section{Goodness-of-fit analysis}

To assess whether the estimated model adequately represents the generative mechanisms of the dynamic network of the patent data at hand, we rely on the goodness-of-fit methods proposed by \citet{Hunter2008b} and simulate 200 networks for each time point $i \in \mathcal{T}$. We then compute the degree distributions for inventors and patents and compare them to the respective observed statistics. For the analysis carried out in this paper, we have 16 time points for which we separately estimated the model. Resulting from this, we also have to carry out the goodness-of-fit assessment for each time point separately. To save space in the supplementary materials, we only show the results for the first and last year in the analysis, namely 2000 and 2015, in Figure \ref{fig:gof}.

\begin{figure}[t!]\centering
	\FloatBarrier
	\includegraphics[trim={0cm 0cm 0cm 0cm},clip,width=0.9\textwidth]{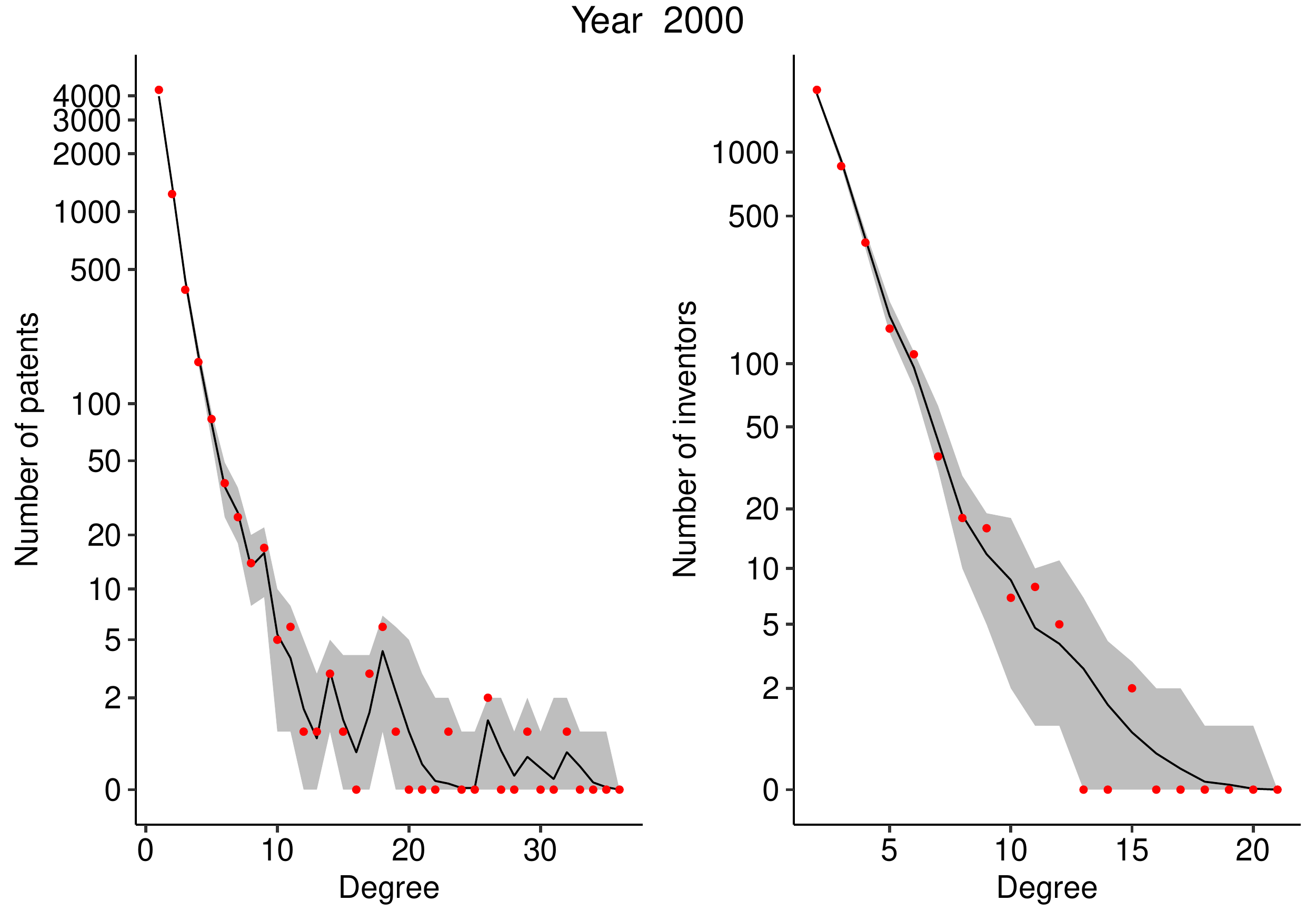}
	\includegraphics[trim={0cm 0cm 0cm 0cm},clip,width=0.9\textwidth]{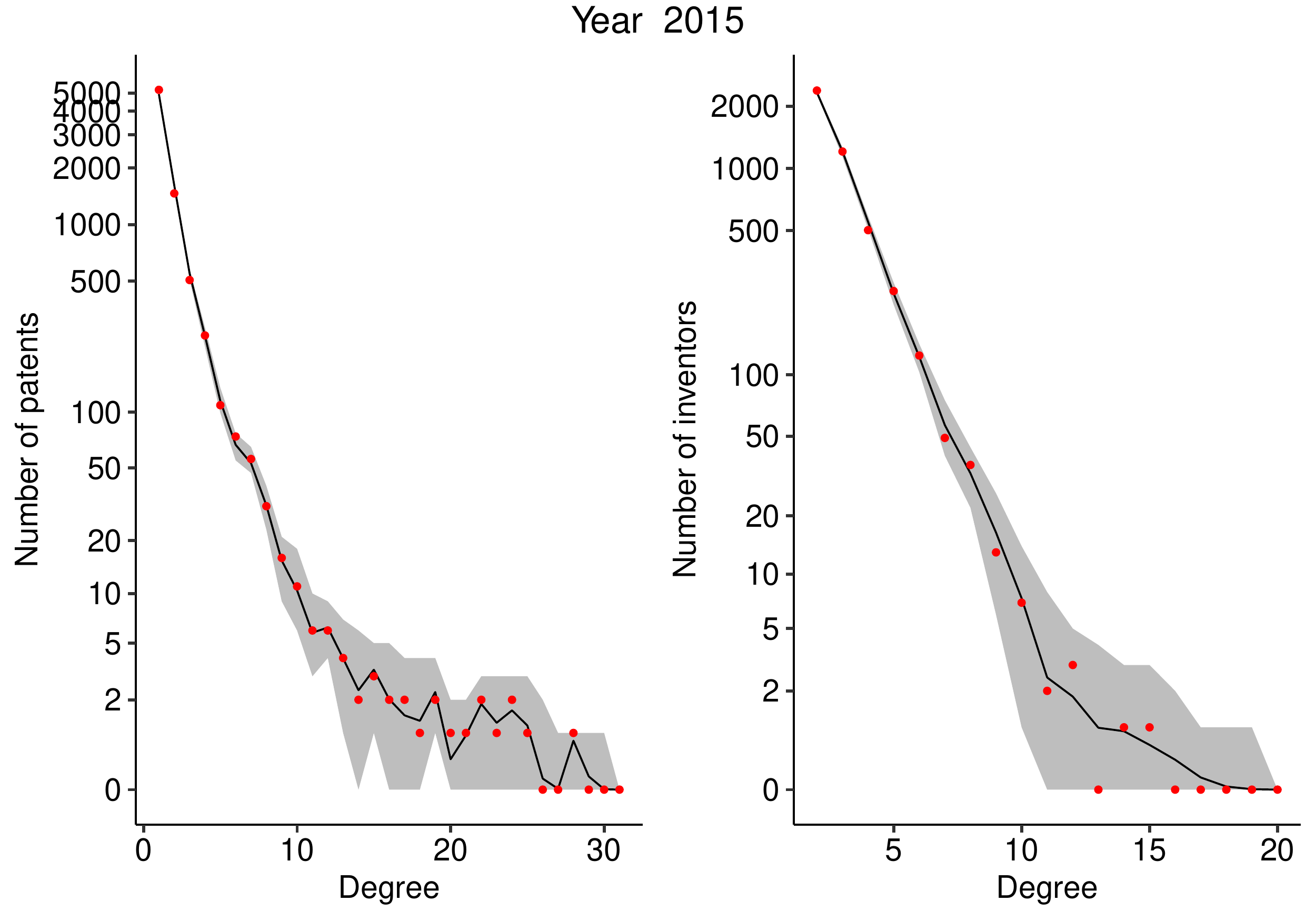}
	\caption{Goodness-of-fit assessment via the degree distributions of inventors and patents on a logarithmic scale in year 2000 and 2015. The red dots show the observed networks, the shades are gives the range for the simulated networks}
	\label{fig:gof}
\end{figure}

\section{MCMC diagnostics}

We also provide the standard MCMC diagnostics to guarantee converged estimates of $\boldsymbol{\theta}$. Similarly to the goodness-of-fit assessment, we limit the respective trace and density plots to the first and last year of the analysis.

\begin{figure}[t!]\centering
	 \centering
    \includegraphics[width=\linewidth]{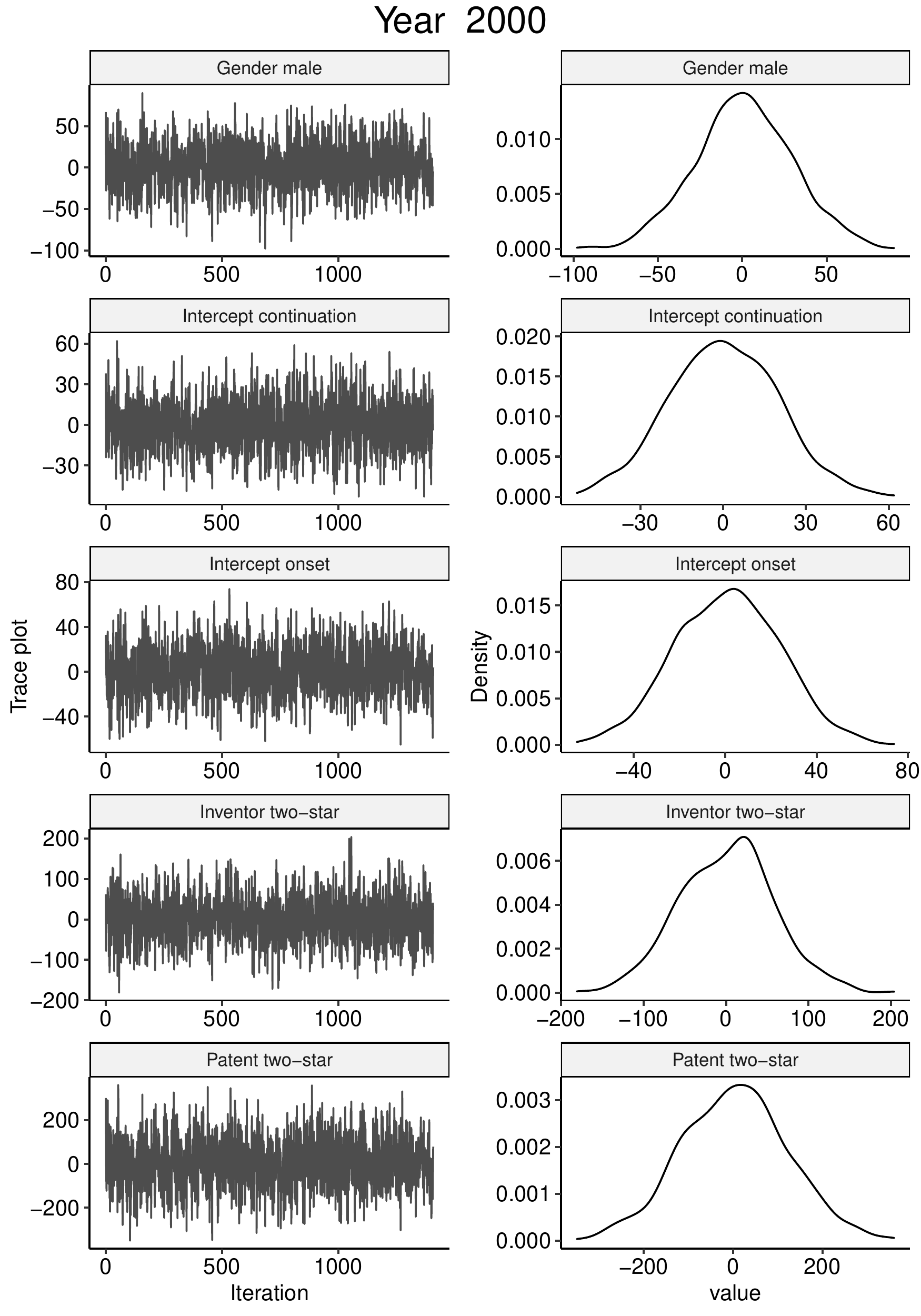}
	\caption{MCMC diagnostics of year 2000 }
	\label{fig:gof}
\end{figure}

\begin{figure}[t!]\centering\ContinuedFloat
	 \centering
    \includegraphics[width=\linewidth]{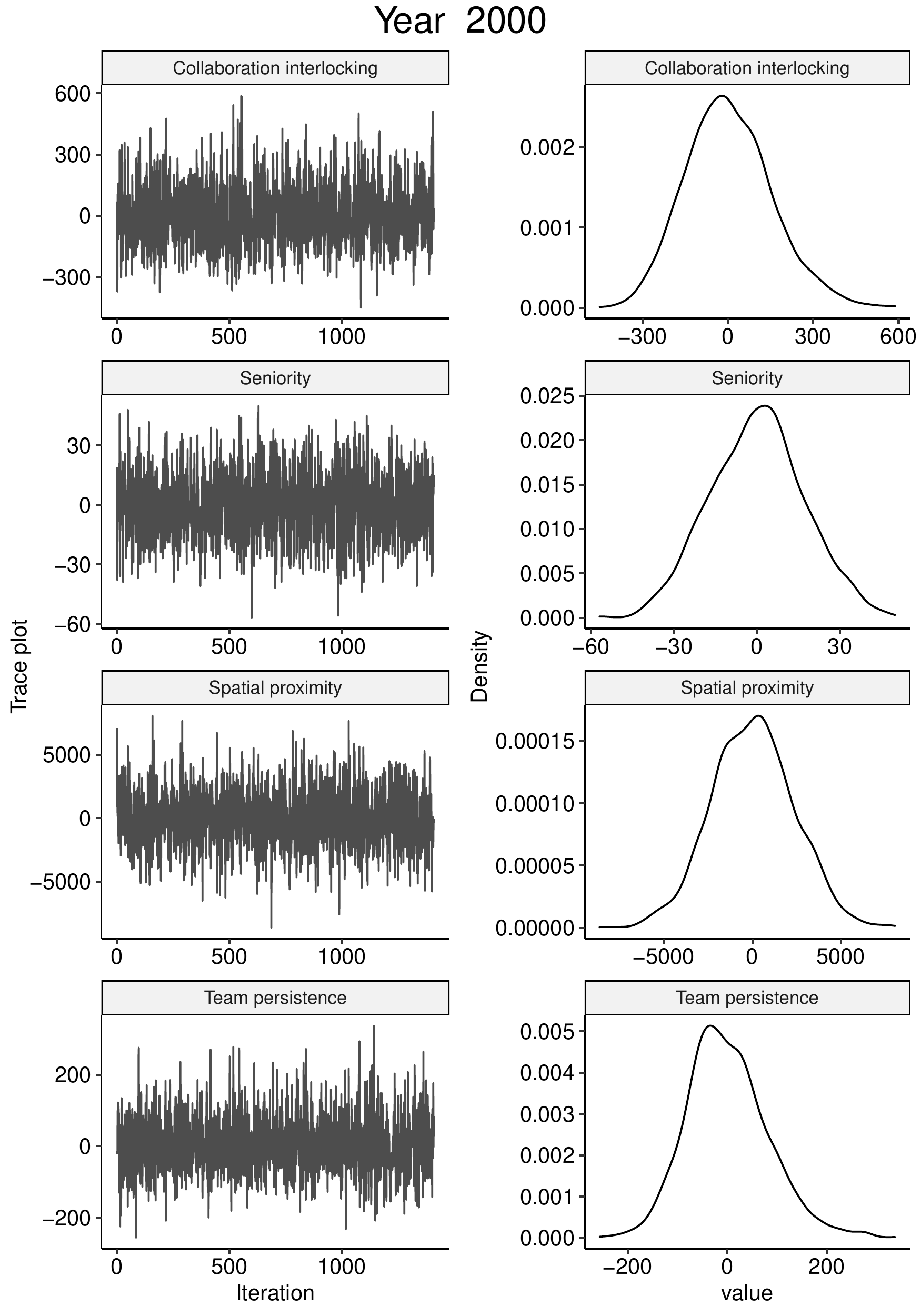}
	\caption{MCMC diagnostics of year 2000 }
	\label{fig:gof}
\end{figure}

\begin{figure}[t!]\centering
	 \centering
    \includegraphics[width=\linewidth]{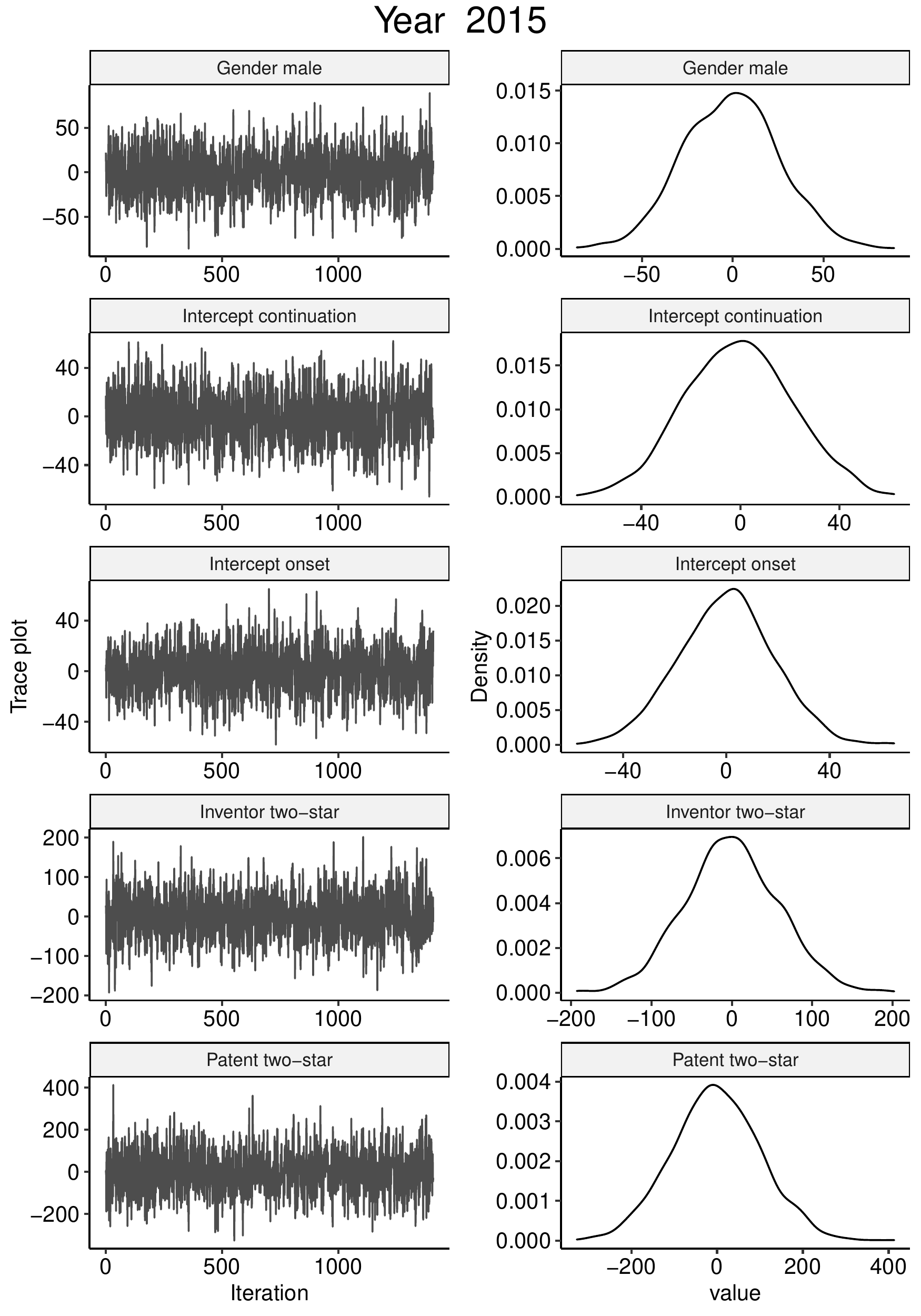}
	\caption{MCMC diagnostics of year 2015 }
	\label{fig:gof}
\end{figure}

\begin{figure}[t!]\centering\ContinuedFloat
	 \centering
    \includegraphics[width=\linewidth]{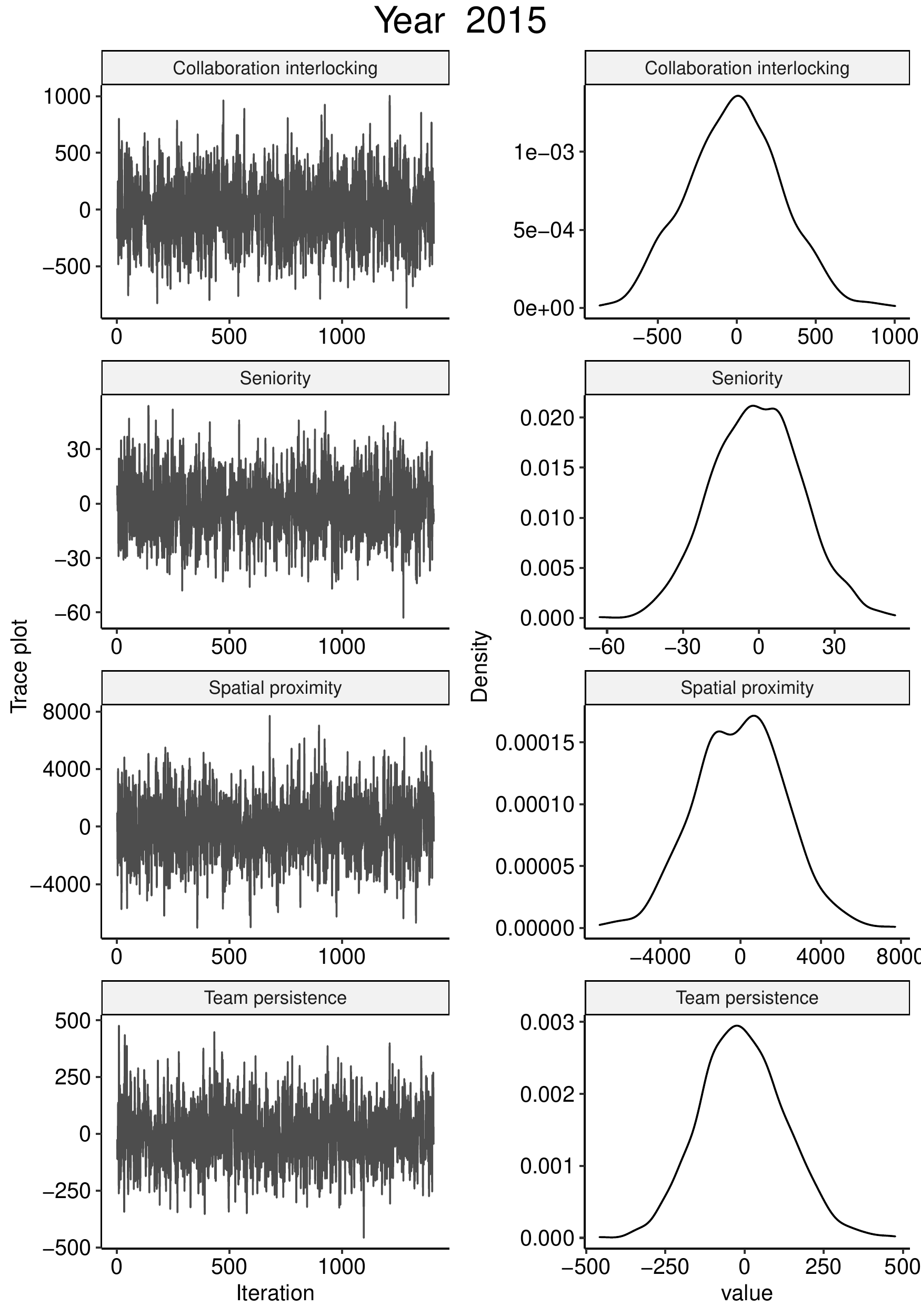}
	\caption{MCMC diagnostics of year 2015 }
	\label{fig:gof}
\end{figure}
\end{document}